\documentclass[preprint,authoryear,12pt]{elsarticle}

\usepackage{graphicx}
\usepackage{amssymb}
\usepackage{float}
\usepackage{url}
\usepackage{amsmath}
\usepackage{placeins}
\usepackage{a4wide}
\usepackage[english]{babel}

\journal{Physica D:  Nonlinear Phenomena}

\begin{document}

\begin{frontmatter}



\title{Noisy dynamic simulations in the presence of symmetry:  data alignment and model reduction}


\author[label1]{Benjamin Sonday}
\author[label1,label2]{Amit Singer}
\author[label4,label1,label3]{Ioannis G. Kevrekidis}

\fntext[label4]{Corresponding author:  yannis@princeton.edu}
\address[label1]{Program in Applied and Computational Mathematics, Princeton University, Princeton, NJ 08544}
\address[label2]{Department of Mathematics, Princeton University, Princeton, NJ 08544}
\address[label3]{Department of Chemical and Biological Engineering, Princeton University, Princeton, NJ 08544, USA}

\begin{abstract}
We process snapshots of trajectories of evolution equations with
intrinsic symmetries, and demonstrate the use of recently developed
eigenvector-based techniques to successfully quotient out the
degrees of freedom associated with the symmetries in the presence of
noise.
Our illustrative examples include a one-dimensional evolutionary
partial differential (the Kuramoto-Sivashinsky) equation with
periodic boundary conditions, as well as a stochastic simulation of
nematic liquid crystals which can be effectively modeled through a
nonlinear Smoluchowski equation on the surface of a sphere.
This is a useful first step towards data mining the
``symmetry-adjusted" ensemble of snapshots in search of an accurate
low-dimensional parametrization (and the associated reduction of the
original dynamical system).
We also demonstrate a technique (``vector diffusion maps") that
combines, in a single formulation, the symmetry removal step and the
dimensionality reduction step.
\end{abstract}

\begin{keyword}
Dimensionality reduction \sep heat kernel \sep local principal
component analysis \sep alignment

\end{keyword}

\end{frontmatter}









\section{Introduction}

High-dimensional dynamical systems are often characterized by low-dimensional long-term dynamic behavior.
Obtaining reduced-dimensionality models consistent with this
behavior is clearly very useful in both analysis and in
computations.
While such model reduction can be based on properties of the
dynamics (e.g. Center-Manifold or Lyapunov-Schmidt reduction, see
\cite{guckholmesbook,lsbook}, or Inertial and Approximate Inertial
Manifolds, see
\cite{jolly1989explicit,aim,aimshort,initmanbook,foias1988inertial,titi1990approximate,foias1989exponential}),
semi-empirical methods based on data-mining are also enjoying
extensive use in applications (e.g. PCA/POD-Galerkin methods, see
\cite{podg1,podg2,podg3,sirisup}).
As nonlinear extensions of Principal Component Analysis are
developed (e.g. techniques like Isomap, Local Linear Embedding,
Laplacian Eigenmaps/Diffusion Maps, etc., see
\cite{isomap,lle,belkin2003,coiffp,coifman_diffusion}), the
necessity of linking these nonlinear data reduction techniques with
dynamic model reduction naturally arises.

When the data set of interest consists of snapshots of trajectories
of dynamical systems with symmetry, ``factoring out" this symmetry
is an established first step (in theory, in computations, as well as
in PCA-based data mining); the use of so-called ``template
functions" in this context has been described by Rowley and
coworkers (e.g. \cite{rowley1,rowley2}, see also
\cite{pres_symm_pod,holmes1998turbulence}).
In this paper we explore the application of recently developed
computational approaches to symmetry removal (``factoring out"
symmetry, ``alignment" of the data) for (noisy) high-dimensional
dynamical system data.
Our illustrative examples include (1) the discretization of a
well-known spatiotemporal pattern-forming partial differential
equation (PDE), the Kuramoto-Sivashinsky Equation (KSE) in one
spatial dimensional and with periodic boundary conditions (with
associated symmetry group $SO(2)$); and (2) a stochastic simulation
of a nonlinear 2D Smoluchowski equation, where the evolution of the
orientational distribution function of an ensemble of nematic liquid
crystals is modeled on the sphere (with associated symmetry group
$SO(3)$).
In both cases noise is present in the data; in the KSE case the
noise is added externally (by us); in the nematic liquid crystal
case the noise comes from the stochastic simulation of a finite
ensemble of representative particles.

The essential step in factoring out the relevant symmetries involves
relating each snapshot in the data to each other snapshot (in
effect, using each snapshot as the ``alignment template" for every
other snapshot); using these pairwise relations to perform a global
alignment can be formulated as an optimization problem that is
fruitfully relaxed to an eigenproblem (hence the term ``eigenvector
method," see \cite{amit,cryo1,cryo2}).

In one of our examples (the KSE) we will also demonstrate the
combination of this ``alignment" with a second, data mining
(dimensionality reduction) step; the combination carries the name of
``vector diffusion maps" (\cite{vdm}) and has potential advantages
over the ``two step" approach (first alignment and then reduction).
The data set corresponding to the snapshots of the dynamical system
is usually modeled as lying on a low dimensional manifold
$\mathcal{M}$. In the presence of a symmetry group $G$ (such as
$SO(2)$ or $SO(3)$), vector diffusion maps provide a natural
framework to organize the data in the quotient space $\mathcal{M} /
G$. The affinities between data points are related to their
correlation when they are optimally aligned, and the information
about the optimal alignment transformation (the group element) is
also encoded in this framework. The advantage of working in the
quotient space $\mathcal{M} / G$ stems from its lower dimensionality
compared to the original manifold $\mathcal{M}$, giving rise to
improved dimensionality reduction, noise robustness, and the need
for less data.

The paper is organized as follows.
In Section \ref{sec:probDesc}, we give an overview of the
``alignment" problem and briefly review template-based methods.
Next, Section \ref{sec:SummaryMethod} summarizes the eigenvector
method and some of its relevant mathematical properties.  Sections
\ref{sec:LCP} and \ref{sec:KSE} are devoted to applying and
comparing template-based approaches and the eigenvector method to
our two prototypical examples. Finally, in Sections \ref{sec:DMAPS}
and \ref{sec:vDMAPS}, we demonstrate the use of two dimensionality
reduction techniques, diffusion maps and vector diffusion maps, on
the modulated traveling wave data of Section \ref{sec:KSE}.

\section{Description of the problem} \label{sec:probDesc}

For physical systems possessing symmetry, there may be several
equivalent realizations of what is effectively the same system state
(whether a steady/stationary state or an ``instance" or ``snapshot"
during a transient simulation); these realizations are related by
some underlying symmetry group.
When such systems with symmetry evolve in time, their dynamics
are equivariant with respect to the appropriate symmetry group.
Consider a function $u(\theta,t)$ on the unit circle evolving
according to some spatially invariant differential operator
$\mathcal{D}$ via an equation of the form
\begin{equation}
u_t = \mathcal{D}(u).
\end{equation}
This equation is equivariant in the sense that
\begin{equation}
\mathcal{D}(S_c[u]) = S_c[\mathcal{D}(u)],
\end{equation}
where $S_c[v](\theta) = v(\theta+c)$ is the \textit{shift operator}
on spatially periodic functions; starting at a particular snapshot,
evolving the dynamics for some time and shifting the final state by
$c$ is the same as the result of shifting the initial snapshot by
$c$ and then evolving the dynamics from the shifted initial
condition (in other words, the differential operator $\mathcal{D}$
commutes with the shift operator $S_c$).

Suppose we take $M$ snapshots of $u$ at $M$ different times, $\{
u(\theta,t_k) \}_{k=1}^M$.
If $u(\theta,t)$ is not
changing its shape, but simply traveling around the unit circle (for example,
when $\mathcal{D}(u) = \omega u_\theta$), we may identify each
snapshot with some angle $\theta \in [0,2 \pi)$.
By rotating each of these snapshots ``back'' by the angle $\theta$
with which it has been identified, we obtain a set of
\textit{identical} system snapshots (thereby removing one degree of
freedom from the evolving system).
%

\begin{figure}
\begin{center} \includegraphics[width=150mm]{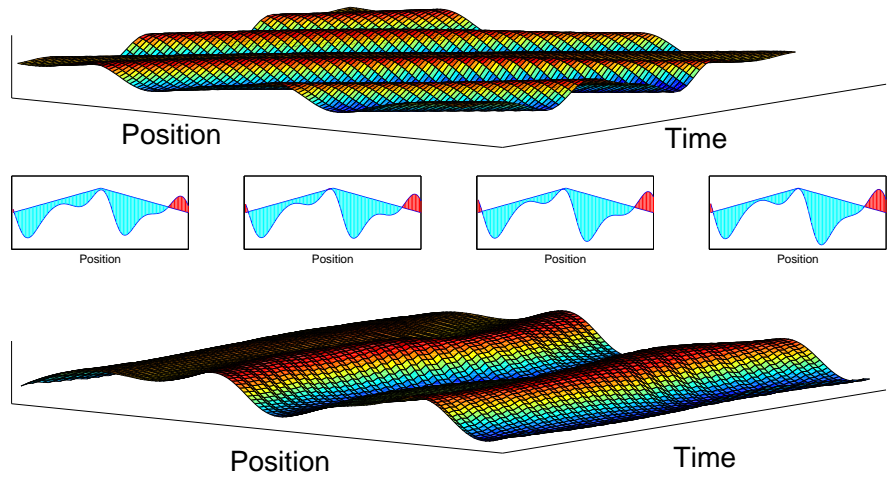}
    \caption{At the top, a representative schematic of a system evolving in space and time.  Here, the domain is periodic in the spatial direction.  Using a triangle shape as a \textit{template} (see text for a discussion of templates), we spatially shift each dynamic snapshot (each time slice of the top figure) to maximally correlate with the triangular template.  Four such maximal correlations are shown in the middle figure, where the shading brings out the difference between the snapshot and the template. At the bottom, we see that after alignment, the dynamics of the set of snapshots appears visually much simpler; the traveling motion is gone, and all that remains is a slight modulation.}
    \label{fig:YannisPlot}
\end{center}
\end{figure}

The removal of this degree of freedom allows us to perform certain
tasks, such as denoising a collection of snapshots through averaging
(in \cite{cryo1,cryo2}, a similar procedure is used on cryo-EM
data), more easily.
In the case where $u(\theta,t)$ is evolving its shape in addition to
traveling (for example, when $\mathcal{D}(u) = \omega u_\theta +
\mathcal{E}(u)$, where $\mathcal{E}(u)$ is some other nonlinear
spatially invariant operator), removing this ``traveling'' degree of
freedom from the simulation can significantly assist our
understanding of the dynamics.
See Figure \ref{fig:YannisPlot} for an illustration.
For instance, when one uses \textit{diffusion maps} to explore
whether the simulation data are intrinsically low-dimensional, and
to find good ``coarse" parametrizations for them (see, e.g.,
\cite{lafonlee,physreve,cecilia,coifman2,frewenchem}), removing the
symmetry results in a more parsimonious description of the dynamics
(an embedding in a lower-dimensional space), which may also be
successfully deduced with far less data.

Now, suppose we have an ensemble of $M$ snapshots, but we do not
know the members of the underlying symmetry group with which each
snapshot is to be identified.
We wish to perform this association of snapshots with symmetry group
elements; in other words we wish to globally ``align" the  $M$ snapshots.
(Here the colloquial expression ``alignment" comes from the simple conceptual example
of rotationally invariant functions on the unit circle; the possible
rotation angles can be ``strung" along a line between $0$ and $2 \pi$.)

Normally, this global alignment (the computation of the symmetry
group element identified with each snapshot) may be accomplished
numerically through the use of a well-chosen {\em template function}
(see Figure \ref{fig:YannisPlot} and, e.g., \cite{rowley1,rowley2}).
For instance, in our running example of snapshots $\{ u(\theta,t_k) \}_{k=1}^M$,
one finds the alignments $\{ \theta_k \}_{k=1}^M$ which align each snapshot
with a template $T(\theta)$ by simply setting
\begin{equation} \label{eq:maxTempl}
 \theta_k = \underset{ c }{\operatorname{argmin}}\,\, \| T(\theta)-u(\theta+c,t_k) \|^2.
\end{equation}
The analogue of equation (\ref{eq:maxTempl}) holds for other
symmetry groups.
This approach will, in general, be
successful when
\begin{itemize}
\item there is little noise in the data;
\item a ``good" template, leading to a clear global minimum, is known ahead of
time; and
\item  this template remains ``good" in the above sense as new data are collected during the system evolution.
\end{itemize}
When there is noise in the data, or when a good template is not known,
``misalignments'' may happen frequently.
Furthermore, as the system evolves, a fixed template may stop being
``good" (that is, giving rise to a clear global minimum in the above
optimization problem).

In this paper we apply a novel spectral algorithm (\cite{amit}) to
solve this problem of global alignment in the presence of symmetry.
In contrast to the method of templates, which compares snapshots one
by one to a fixed ``template function'' (producing $M$ pieces of
information), the eigenvector method compares {\em all} snapshots to
{\em all other} snapshots \textit{pairwise}, in essence treating
\textit{every} snapshot as a template (and thereby exploiting a
greater amount of information, namely $M(M-1)/2$ pieces).
Even though many of these pairwise comparisons
may be inaccurate due to noise inherent in the snapshots, consistency
relationships among these pairwise alignments can be used to gain a
sense of the overall, {\em global} alignment.
A slight modification of this algorithm known as \textit{vector
diffusion maps} (\cite{vdm}) allows for the situation in which the
snapshots differ not only by a symmetry group element (and noise),
but also because there is a systematic change in the snapshots due
to the underlying dynamic evolution.
Both algorithms are fast, simple, and (as we will demonstrate) more robust to noise than
their corresponding template-based approaches.

The ``eigenvector algorithm" will be illustrated  through two
prototypical examples.  The first involves the evolution of orientational distribution functions of
nematic liquid crystal polymers; the distributions are functions on
the sphere, and we take the associated symmetry group to be $SO(3)$.  The
second involves spatiotemporally traveling/modulating  waves of the
Kuramoto-Sivashinsky equation (KSE); these are functions on the unit circle with
periodic boundary conditions, and we take the associated
symmetry group to be $SO(2)$.
Additionally, for the case of the KSE waves, we demonstrate the use of vector diffusion maps to
(all in a single step), remove the underlying symmetry {\em and} capture the low-dimensionality of the
underlying dynamics (the residual dynamics of modulation after the ``traveling'' symmetry has been removed).



\section{The eigenvector alignment method} \label{sec:SummaryMethod}

In its most general form, the eigenvector alignment method
(\cite{amit}) can be summarized as follows.
Consider an ensemble of $M$ snapshots which are identical, except
for the action of some underlying symmetry group $G$ (such as
spatially periodic translation) and perhaps some noise.
We wish to know the group elements $\{ g_i
\}_{i=1}^M \in G$ with which the $M$ snapshots may be identified; this
will give us information which can be used to, for instance,
ascertain what ``rotation'' to perform to make a particular snapshot
equivalent to another (i.e. to ``align" the two snapshots).
Specifically, if we identify snapshots $i$
and $j$ with group elements $g_i$ and $g_j$, then rotation of
snapshot $i$ by $g_{ij} \equiv g_j g_i^{-1}$ should make it
identical to snapshot $j$.
In our simple illustration of periodic
functions $u(\theta,t)$ traveling around the circle with speed $\omega$
(dynamics $u_t = \omega u_\theta$), the symmetry group elements are angles
modulo $2 \pi$ (the group is $SO(2)$).
Each snapshot $u(\theta,t_i)$ can, in principle, be
identified with some angle $\theta_i$.
Snapshot $i$ may be made
equivalent to snapshot $j$ after a (say, systematically counter-clockwise) rotation of snapshot $i$ by
$\theta_{ij} \equiv \theta_j-\theta_i$.

When the snapshots are noise-free, obtaining the $\{ g_i \}_{i=1}^M$
may be done easily as follows.
Choose one \textit{base snapshot},
or ``template,'' say snapshot $i$.
For this snapshot $i$, choose a particular random assignment $g_i$.
For each remaining snapshot $j$, find the
$g_{ij} \in G$ which rotates snapshot $i$ to be identical to
snapshot $j$, and then set $g_j = g_{ij} g_i$.
Alignments between
any two snapshots $p$ and $q$ can then be computed as $g_{pq} =
g_{ip}^{-1} g_{iq} $.
In the example of angles modulo $2 \pi$, this
means choosing some base $\theta_i$ for snapshot $i$, then setting
$\theta_j$ (for each of the remaining snapshots) to be $\theta_j =
\theta_{ij} + \theta_i$ (where $\theta_{ij}$ is the angle which
rotates snapshot $i$ to be identical to snapshot $j$).
Alignments
between any two snapshots $p$ and $q$ can then be computed as
$\theta_{pq} = \theta_{iq}-\theta_{ip}$ (to get from snapshot $p$ to
$q$, rotate snapshot $p$ back to snapshot $i$, then rotate $i$ to
$q$).

Because the method above relies on using only \textit{a single}
template, it may well not be robust to noise; obtaining the $g_j$
may not work well because many of the $\{ g_{ij} \}_{j=1}^M$ will be
computed incorrectly.
The {\em eigenvector method} instead has the user compute
all $\{ g_{ij} \}_{i,j=1}^M$ (in essence, treating \textit{every}
snapshot as a template); it then looks for {\em consistency} along these
pairwise alignments to assign the global alignments $\{ g_i
\}_{i=1}^M$.
The main idea is as follows:  if $g_{ij}$, $g_{jk}$, and
$g_{ik}$ are accurately measured, we also expect, for example, that
\begin{equation}
 g_{ik} = g_{ij} g_{jk}, \label{eq:tripCons}
\end{equation}
a condition known as the {\em triplet consistency relation}.
In our example of angles modulo $2 \pi$, this simply says that,
regardless of whether snapshot $i$ or $j$ are used as the template,
the angle between snapshot $i$ and snapshot $k$ should be the same
no matter if it is measured directly ($\theta_{ik}$) or inferred
($\theta_{ij}+\theta_{jk}$).
Analogously, we also expect
``higher-order'' consistency relations of the form
\begin{equation}
 g_{il} = g_{ij} g_{jk} g_{kl}. \label{eq:quadCons}
\end{equation}
Since many of the measurements of $g_{ij}$ may be inaccurate,
equations (\ref{eq:tripCons}), (\ref{eq:quadCons}), and their
high-order forms will often be violated; however, one can still hope
to assign the $g_i$ in some sort of globally optimally consistent
way.

Initially, one may attempt to assign the $g_i$ so that as many
pairwise measurements $g_{ij}$ as possible are satisfied to within
some tolerance.
Unfortunately, for even a moderate number of group
elements $M$, it is computationally intractable to find the
assignment of the $g_i$ which maximizes the number of them which are
satisfied (to within some tolerance).
This is a non-convex optimization problem in a very high dimensional
space.
As we discuss now on the example of angles modulo $2 \pi$, a {\em
relaxation} of the problem to a quadratic (and therefore convex)
form has been proposed (\cite{amit}).
The only requirement is for the symmetry group $G$ to have a compact
real/complex form.
The relaxation makes the optimization problem more tractable, but it
also allows for the ``solution" $g_i$ to include elements not
necessarily in $G$ (we will explain this and show how it can be
rectified below).

Again, consider the problem of angles modulo $2 \pi$.
This group
has a compact complex representation given by mapping $\theta_i$ to
$e^{i \theta_i}$.
Measurements of $\theta_{ij}$, which are (noisy)
measurements of $\theta_j-\theta_i$, are represented similarly as
$e^{i \theta_{ij}}$.
At first, one might wish to formulate the problem so as to assign
the global alignments $\theta_i$ in order to maximize the number of
pairwise measurements which hold true to within some tolerance
$\mbox{tol}$, for instance
\begin{equation}
\underset{\{\theta_i\}}{\operatorname{argmax}}\,\,\# \{ (i,j):\,\, - \mbox{tol} \leq \theta_j-\theta_i-\theta_{ij} \,\,(\mbox{mod}\,\, 2 \pi) \leq \mbox{tol}  \}.
\end{equation}
This problem becomes quickly computationally intractable for large $M$, even
after a reformulation to the form
\begin{equation}
\underset{\{\theta_i\}}{\operatorname{argmin}}\,\, \sum_{ (i,j)} f\left[\theta_j-\theta_i-\theta_{ij} \,\,(\mbox{mod}\,\, 2 \pi)\right],
\end{equation}
where $f$ is some smooth periodic penalty function.

Instead, the problem is relaxed as follows:  the measurements
$\theta_{ij}$ are inserted into a matrix $\mathbf{H}$ so that
$\mathbf{H}_{ij} = e^{-i \theta_{ij}}$.
We now consider maximizing
the following quantity:
\begin{equation}
 \underset{\{\theta_i\}}{\operatorname{argmax}}\,\, \sum_{i,j}^M e^{-i \theta_i} \mathbf{H}_{ij} e^{i \theta_j}. \label{eq:maxTheta}
\end{equation}
When the $\theta_i$ are correctly assigned, each ``good''
measurement of $\mathbf{H}_{ij}$ contributes close to $1$ in the sum
and each ``bad'' measurement contributes, on average, $0$ to the sum
({\em since the errors are assumed to be uniformly randomly distributed},
see \cite{amit}).
Therefore, the maximization of the expression
(\ref{eq:maxTheta}) is likely to produce, in some sense, maximally
consistent assignments of the $\theta_i$.
To make the problem even
more tractable, it is further relaxed to a quadratic form
(general complex numbers, as opposed to complex numbers on the unit circle only) which can
be easily solved with power iteration:
\begin{equation}
 \underset{\{z_i\} \in \mathbb{C}, \sum |z_i|^2 = M}{\operatorname{argmax}}\,\, \sum_{i,j}^M z_i^* \mathbf{H}_{ij} z_j. \label{eq:maxZ}
\end{equation}
Maximizing the expression (\ref{eq:maxZ}) amounts to finding the largest
eigenvector $v$ of the Hermitian matrix $\mathbf{H}$.
The components of the largest eigenvector $v$ are not necessarily of
unit length, but after normalization, one can define the {\em estimated}
angles by
\begin{equation}
 e^{i \theta_i} = \frac{v(i)}{|v(i)|}.
\end{equation}

It is interesting to note that the error of the assignments
$\theta_i$ can be estimated by looking at the eigenvalue spectrum of
$\mathbf{H}$.
Consider, for instance, the correlation $\rho$
between the eigenvector $v$ and the vector $z$ of true angles as a
measurement of ``goodness of fit''; this is given as
\begin{equation}
 \rho = \left|\frac{1}{\sqrt{M}} \sum_{i=1}^M e^{-i \theta_i} v(i) \right| = |\langle z,v \rangle|. \label{eq:rho}
\end{equation}
Under certain assumptions about the type of noise in the problem, one can
show that
\begin{equation}
 \left| \langle z,v \rangle \right|^2 \geq \frac{\lambda_H-\lambda_R}{Mp},
\end{equation}
where $M$ is as above, and $p$ is a quantity related to how likely
``good'' measurements are (see \cite{amit} for details).
Here $\lambda_H$ is the leading eigenvector of the matrix
$\mathbf{H}$; if the (random) matrix $\mathbf{H}$ has a number of
properties (again, see \cite{amit,feral}) its eigenvalue
distribution will include a semicircle, and the right edge of this
semicircle will be the quantity $\lambda_R$.
Furthermore,
\begin{equation}
 \lambda_R \approx 2 \sqrt{M(1-p^2)}
\end{equation}
and
\begin{equation} \label{eq:lambda_Hexp}
 \mathbb{E}\left[\lambda_H\right] \approx Mp + \frac{1-p^2}{p},
\end{equation}
where equation (\ref{eq:lambda_Hexp}) is valid whenever $p > 1 /
\sqrt{M} $ and the variance in the quantity $\lambda_H$ increases as
$p$ decreases (\cite{amit,feral}).
%

%
%
Although the noise model presented in \cite{amit} is different than
the noise in our problems, equation (\ref{eq:rho}) holds regardless, and we
still expect the alignment error to decrease as both $M$ and $p$ grow
(more data/pairwise comparisons and higher quality measurements, respectively, will lead to
a better recovery of the global alignments).

We also note that \cite{feral} requires the noise in every entry of the matrix
$\mathbf{H}$ to be {\em independent}.
This is not necessarily true in our examples.
It is likely that ``good'' and ``bad'' measurements are not random,
but rather, correlated; having independent entries requires $M^2$
sources of randomness, and clearly, for large enough $M$, this will
cease to be true because the ``amount'' of randomness scales only as
$M$, the number of snapshots.
For large $M$ this argument can rationalize why some eigenvalues
(with magnitude of $O(M)$) may appear \textit{outside} the
theoretically expected semicircle (see, e.g., Figures
\ref{fig:SHeigs} and \ref{fig:eval_KSE}).
This phenomenon is investigated in \cite{Xiuyuan}.

\section{The first illustrative example: orientational distributions of nematic liquid crystal
polymers} \label{sec:LCP}

Symmetry often plays an important role in systems with spontaneous
spatiotemporal pattern formation; such systems, typically modeled
through partial differential equations, arise naturally in modeling
reaction-diffusion and/or flow (\cite{partexamp1}), but also
nonlinear optics (\cite{partexamp2}) and Bose-Einstein condensates
(\cite{partexamp3}).
If the computational models are in the form of stochastically interacting particles,
the finite number of the simulated particles and the stochasticity of their evolution
naturally gives rise to noise in the recorded time series (and we know that the
fewer the particles, the ``larger" in some sense the noise will be).
To illustrate this, and to show how to factor out symmetries at the
``macroscopic" level while working with a ``microscopic,'' particle
based, noisy simulation, we chose an illustrative example for which
good models exist at both the particle- and the continuum levels.
The system in question is the evolution of the single particle
orientational probability distribution function in the case of
nematic liquid crystals; a closed equation that very successfully
approximates this evolution is a Smoluchowski equation (\cite{LCP}).
An alternative description of the dynamics comes in the form of
coupled stochastic differential equations which model the
interactions of a (large but) finite number of nematic liquid
crystal polymer molecules; one hopes that, for a large enough number of
simulated interacting particles, the
computed evolution of their collective orientational probability distribution
approximates the trajectories of the (mesoscopic) Smoluchowski equation.

It is well known (and can be seen from the form of equation \ref{eq:Smol} below)
that the evolution of the orientational probability distribution
is characterized by equivariance: rotating the initial distribution
on the unit sphere and evolving commutes with evolving for the same amount
of time and then rotating the final distribution.
This implies that experiments (or simulations) differing by some
(unknown) mesoscopic rotation of the entire initial distribution
should, in effect, produce the same results (modulo the effects of
noise).


\subsection{System setup}
Liquid crystalline polymers (LCPs) are large molecules which contain
long rigid segments.
Groups of LCPs are capable of displaying rich
behavior including high modulus in the solid phase, low viscosity in
the melt, and many other interesting and/or desirable physical
properties.
Each LCP can be thought of as a ``needle,'' whose orientation may be
described as a pair of antipodal points $\pm \mathbf{w}$ (the
``tips'' of the needle) on the unit sphere; as the number of LCPs in
a group becomes large, the evolution of the single-particle
orientational probability distribution function $\psi(\mathbf{u})$
of the group is accurately described by the Smoluchowski equation
\begin{equation} \label{eq:Smol}
 \frac{\partial \psi(\mathbf{u})}{\partial t} = D \frac{\partial}{\partial \mathbf{u}} \cdot \left[\frac{\partial \psi(\mathbf{u})}{\partial \mathbf{u}} + \psi(\mathbf{u}) \frac{\partial}{\partial \mathbf{u}} \left( \frac{V[\psi,\mathbf{u}]}{kT} \right) \right].
\end{equation}
Here, $\mathbf{u}$ is a unit vector describing orientation,
$\partial / \partial \mathbf{u}$ is the gradient operator restricted
to the unit sphere, k is Boltzmann's constant, T is the absolute
temperature, D is the rotational diffusivity (here set to $1$), and
$V[\psi,\mathbf{u}]$ is a nematic potential (a free energy taking
into account excluded volume effects).
For our simulations, we use the Maier-Saupe potential (see, e.g.
\cite{maiersaupe})
\begin{equation}
 V[\psi,\mathbf{u}] = -\frac{3}{2} U \mathbf{u}\mathbf{u}:\mathbf{S},
\end{equation}
where $\mathbf{S}=\langle \mathbf{u} \mathbf{u} \rangle -
\frac{1}{3} \mathbf{I}$ is the \textit{tensor} order parameter.
The
parameter $U$ (the intensity of the nematic potential) can be
thought of as proportional to the concentration of the LCP ``rods".
If $\lambda$ is the eigenvalue of $\mathbf{S}$ with the largest
magnitude, the so-called \textit{scalar order parameter} $S$ is
given by $S=3 \lambda / 2$ (\cite{LCP}).
Writing equation (\ref{eq:Smol}) as $\partial
\psi(\mathbf{u})/\partial t = \mathcal{D}(\psi(\mathbf{u}))$, the
Smoluchowski equation is equivariant in the sense that
$\mathcal{D}(\psi(\mathbf{R}\mathbf{u})) = \mathbf{R}
\mathcal{D}(\psi(\mathbf{u}))$, where $\mathbf{R}$ is a member of
$SO(3)$.

Computationally, the evolution of the distribution function can be
simulated as a large set of coupled stochastic differential equations.
One simply represents the distribution $\psi(\mathbf{u})$ as a
collection of $N$ representative individual LCPs, and then computes
their trajectories $\{ \pm \mathbf{w}_i(t) \}_{i=1}^N$ (here, $\{
\pm \mathbf{w}_i(t) \}_{i=1}^N$ are vectors on the surface of the
sphere, and the ``$\pm$'' is because each LCP is really a rod with
identical ``top'' and antidiametric ``bottom'').
Initializing a distribution $\psi_0(\mathbf{u})$ with $N$ particles
may be done with the Metropolis-Hastings algorithm (see
\cite{metropolis} or the Appendix), and as $N$ goes to infinity,
this initialization converges in measure to $\psi_0(\mathbf{u})$.
Using the $N$ particle trajectories, ensemble averages $\langle
f(\mathbf{u}(t)) \rangle$ at any time $t$ may be evaluated as
$\frac{1}{2N}\sum_{i=1}^{N} f(\mathbf{w}_i(t))+f(-\mathbf{w}_i(t))$
(where here, again, we have a ``$-$'' due to the fact that each LCP
has a top and a bottom).
The distribution $\psi_t(\mathbf{u})$ at time $t$ may also be
reconstructed by a variety of techniques; here we choose to do the
reconstruction by evaluating ensemble averages of the form
$\frac{1}{2N}\sum_{i=1}^N Y_l^{m*}(\mathbf{w}_i(t)) +
Y_l^{m*}(-\mathbf{w}_i(t))$ (these $Y_l^{m*}$ are the spherical
harmonics coefficients of $\psi_t(\mathbf{u})$, see Section
\ref{sec:SH}).
The explicit Euler-Maruyama integration of each individual
(stochastic) trajectory takes the form
\begin{equation}\label{eq:Euler}
\mathbf{w}_i(t+\Delta t) = \frac{\mathbf{w}_i(t)-\frac{D}{kT}\frac{\partial V}{\partial \mathbf{u}}  \Delta t + \sqrt{2 D \Delta \mathbf{b}} }{\left|\left|\mathbf{w}_i(t)-\frac{D}{kT}\frac{\partial V}{\partial \mathbf{u}}  \Delta t + \sqrt{2 D \Delta \mathbf{b}}\right|\right|}.
\end{equation}
By using different numbers for $N$, the errors in the initialization
of $\psi_0(\mathbf{u})$, the computations of the $\langle
f(\mathbf{u}(t)) \rangle$, and the reconstruction (from the particles) of
$\psi_t(\mathbf{u})$ can be controlled, since they scale as
$1/\sqrt{N}$.

The evolution of the Smoluchowski equation is equivariant under the
group $SO(3)$; rotating a given orientational probability distribution
by some element $\mathbf{R}$ of $SO(3)$ and evolving is the same
as evolving first, and then rotating the result by the same group element.
In an SDE reformulation of the problem, an orientational probability distribution
is represented by $N$ particles.
For purposes of computational exploration of its evolution,
a particular ensemble of $N$ particles is
equivalent to any other ensemble in which each of the $N$ particles
$\mathbf{w}_i$ is rotated by the same element of the $SO(3)$ group,
$\mathbf{w}_i \rightarrow \mathbf{R}\mathbf{w}_i$.
Furthermore, due
to the randomness of the Metropolis-Hastings algorithm, each
initialization of $\psi_0(\mathbf{u})$ leads to a different
initial ensemble
of $N$ particles (which will accurately represent $\psi_0(\mathbf{u})$ as
$N$ goes to infinity, but which represent $\psi_0(\mathbf{u})$
noisily for finite $N$).
Thus, in the limit of infinite $N$, a
particular ensemble of $N$ particles initialized consistently with a particular
initial probability distribution
$\psi_0(\mathbf{u})$ is equivalent to another ensemble {\em initialized consistently
with $\psi_0(\mathbf{R}\mathbf{u})$}: the original
distribution, but rotated by a member of $\mathbf{R}$ of the $SO(3)$
group.
For finite $N$, there is noise, and these two ensembles of
$N$ particles are only \textit{approximately} the same after
rotation by a member of $SO(3)$.
Finding this corresponding member
$\mathbf{R}$ of $SO(3)$ becomes increasingly difficult as $N$ gets
smaller.

Suppose we are given a set of $M$ LCP ensembles, each initialized with
$N$ particles, each consistently with $\psi_0(\mathbf{R}_i\mathbf{u})$ for
some unknown rotation $\{ \mathbf{R}_i \}_{i=1}^{M} \in SO(3)$;
and let us evolve each of these ensembles for some fixed time $T$.
The result is a set of $M$
ensembles of $N$ particles which should be \textit{approximately} the same
after each is rotated by $ \mathbf{R}_i^{-1} = \mathbf{R}_i^T \in
SO(3)$ (the difference is due to the finiteness of $N$).
We wish to be able to consistently determine the unknown members $ \{
\mathbf{R}_i \}_{i=1}^{M} $ so that we know how to relate each
ensemble of $N$ particles to each other ensemble.
When $N$ is small
(equivalently, when the ``noise'' is large), misalignments are bound to
occur frequently.
Therefore, as before, we expect the eigenvector alignment method to
outperform a method based simply on a fixed template function.

\subsection{Consistent initialization of LCP distributions}

In order to compare the performance of the eigenvector alignment
method with that of the classic template method, we must first
generate appropriate data.
For chosen $M$ (number of ensembles) and $N$ (number of particles),
this can be accomplished by first generating $M$ random members $\{
\mathbf{R}_i \}_{i=1}^{M}$ of $SO(3)$, and then initializing $M$
ensembles of $N$ particles according to the distributions
$\psi_0(\mathbf{R}_i\mathbf{u})$ via the (random)
Metropolis-Hastings algorithm.

\begin{figure}
\begin{center} \includegraphics[width=125mm]{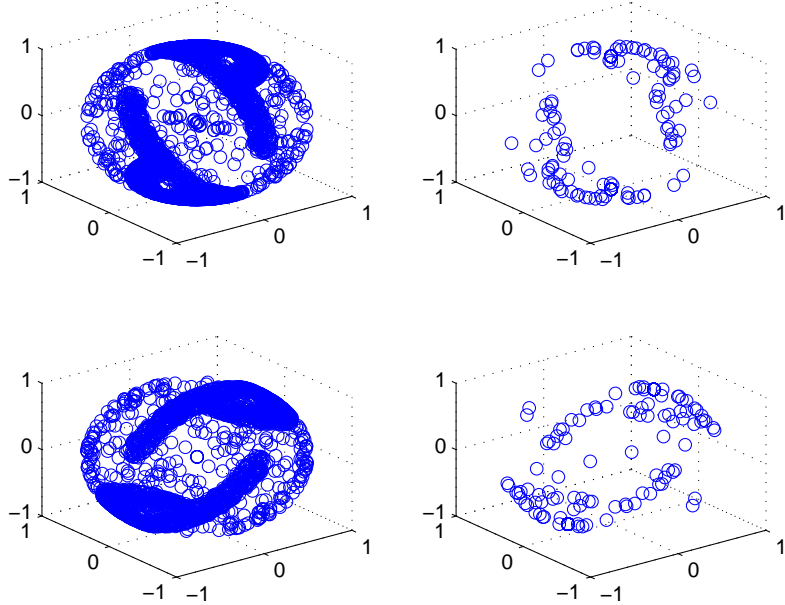} \end{center}
    \caption{Initial distributions of four LCP ensembles from two different rotation matrices (top and bottom
rows, $\mathbf{R}_1\mathbf{u}$ and $\mathbf{R}_2\mathbf{u}$), and
with two different numbers of points ($N=5000$ left, $N=50$ right).
Shown are both the ``top'' and ``bottom'' (interchangeable) of each
LCP particle.}
    \label{fig:init_dist_LCP}
\end{figure}

\begin{figure}
\begin{center} \includegraphics[width=125mm]{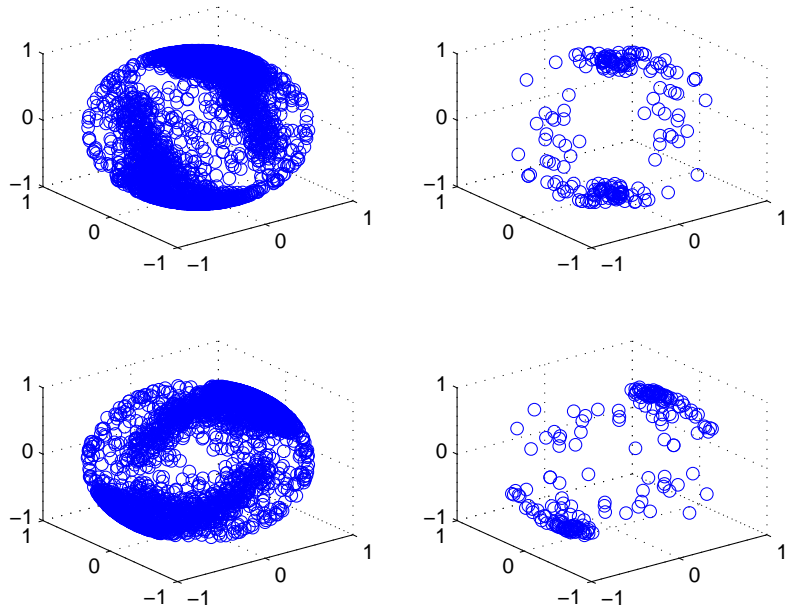} \end{center}
    \caption{The final distributions after integration by $T = .1$s of the LCP ensembles initialized as in Figure \ref{fig:init_dist_LCP}.  Shown are both the ``top'' and ``bottom''
(interchangeable) of each LCP particle.}
    \label{fig:final_dist_LCP}
\end{figure}

This is illustrated through four plots in Figure
\ref{fig:init_dist_LCP}; here, we have plotted both the ``top'' and
``bottom'' (which are interchangeable) of each of the $N$ LCP
particles.
For two random rotation matrices
$\mathbf{R}_1$ and $\mathbf{R}_2$, and for both $N=50$ and $N=5000$,
we show initializations with respect to the probability distribution functions
$\psi_0(\mathbf{R}_1\mathbf{u})$ and
$\psi_0(\mathbf{R}_2\mathbf{u})$.
Here, we selected and initial probability distribution which
resembles a ``P'' shape (along with its reflection through the
origin).  It is given by
\begin{equation}
\psi_0(\mathbf{u}=(x,y,z)) = \frac{1}{\mbox{Norm}} \left\{
     \begin{array}{lr}
       50/51 & z \ge 0 \mbox{ and } |x| \le 0.1 \mbox{ and } y \le 1 / \sqrt{2} \\
       50/51 & z \le 0 \mbox{ and } |x| \le 0.1 \mbox{ and } y \ge -1 / \sqrt{2} \\
       50/51 & z,y \ge 0 \mbox{ and } 0.4 \le \sqrt{x^2 + y^2} \le 0.6 \\
       50/51 & z,y \le 0 \mbox{ and } 0.4 \le \sqrt{x^2 + y^2} \le 0.6 \\
       1/51 & \mbox{else}
     \end{array}
   \right. ,
\end{equation}
where $\mbox{Norm}$ is some normalization so that $\psi_0$
integrates to $1$.
We subsequently evolved these four ensembles for a fixed amount of time $T$
using the algorithm (\ref{eq:Euler}).
The resulting ensembles are
shown in Figure \ref{fig:final_dist_LCP}.

In the numerical experiments to follow, we choose various values of
both $M$ and $N$, thereby generating $M$ different ensembles of $N$
particles corresponding to $\psi_0(\mathbf{R}_1\mathbf{u})$,
$\psi_0(\mathbf{R}_2\mathbf{u})$, $\hdots$,
$\psi_0(\mathbf{R}_{M}\mathbf{u})$.
We then integrated each ensemble for a fixed amount of time $T$
using the Euler-Maruyama scheme (see equation (\ref{eq:Euler})),
obtaining $M$ distributions corresponding to $\psi_T(\mathbf{R}_1
\mathbf{u})$, $\psi_T(\mathbf{R}_2 \mathbf{u})$, $\hdots$,
$\psi_T(\mathbf{R}_{M} \mathbf{u})$.
These distributions differ by (a) a rotation;  (b) the particular
consistent initialization of the $N$ particles; and (c) the
particular (stochastic) particle sample paths computed through the
Euler-Maruyama integration.
%

Given only the noisy particle distributions obtained at time $T$, we wish to determine the
$M$ unknown rotation matrices
$\mathbf{R}_1,\mathbf{R}_2,\hdots,\mathbf{R}_{M}$.
When $N$ is small, the noise (which scales as $1/\sqrt{N}$) makes
this particularly challenging.
\subsection{Alignment of LCP distributions} \label{sec:SH}
Pairwise alignment was performed by both the template method
(alignment of each ensemble member with a fixed template) and by the
eigenvector method (alignment of each ensemble member with each other
ensemble member).
In our work, we utilized the
spherical harmonics components of the orientational distribution functions
(computed based on the particle states)
to perform pairwise alignment of every pair of ensembles of $N$ representative
particles.
Akin to a Fourier basis on the sphere, spherical
harmonics take into account not only lower-order information such as
the center of mass of the distribution  (the first three nontrivial spherical harmonics),
but also its higher-order moments.
Additionally, the leading spherical
harmonics coefficients can be used to quickly compare functions and rotated
versions of these functions on the sphere (see below), so they are
useful for finding optimal pairwise alignments (required by the
eigenvector alignment method).

To align two ensembles of $N$ particles, we first approximated computationally the
leading coefficients of the spherical harmonics expansion of both particle
distributions.
Let the spherical harmonics expansion of the first
distribution be (approximately) given by
\begin{equation}\label{eq:SHint1}
 f(\theta,\phi) = \sum_{l=0}^{l_{\mbox{max}}} \sum_{m=-l}^l f_l^m Y_l^m(\theta,\phi).
\end{equation}
Here, $f_l^m$ is computed as an integral over the surface of the
sphere $\Omega$ via
\begin{equation}\label{eq:SHint2}
 f_l^m = \int_\Omega f(\theta,\phi) Y_l^{m*}(\theta,\phi) d\Omega;
\end{equation}
by representing the particles as delta functions, equation
(\ref{eq:SHint2}) is approximated as
\begin{equation}
f_l^m = \frac{1}{2N} \sum_{i=1}^N Y_l^{m*}(\theta_i,\phi_i) +
Y_l^{m*}(\pi-\theta_i,\phi_i+\pi),
\end{equation}
where $\theta_i$ and $\phi_i$ are the $(\theta,\phi)$ spherical
coordinates of the $i$th particle's orientation vector
$\mathbf{w}_i$ in the distribution (and we include
$(\pi-\theta_i,\phi_i+\pi)$, of course, because each LCP has a top
and bottom which are interchangeable).
It is clear that only the even spherical harmonics coefficients
survive; for the odd ones, $Y_l^{m*}(\theta_i,\phi_i)$ $+$
$Y_l^{m*}(\pi-\theta_i,\phi_i+\pi)$ equals zero.
Similarly, second distribution $g(\theta,\phi)$ may be approximately
described by its coefficients $g_l^m$.

The squared $L^2$ difference $\mathbf{e}(f,g)$ between the two
functions can then be approximated as
\begin{equation}
 \mathbf{e}(f,g) = \sum_{l=0}^{l_{\mbox{max}}} \sum_{m=-l}^l ||f_l^m-g_l^m||^2.
\end{equation}
Once the spherical harmonics expansion of a function $h(\mathbf{u})$
is known, the spherical harmonics expansion of $h_{\mathbf{R}}
\equiv h(\mathbf{R} \mathbf{u})$ can be computed quickly; therefore,
it is only necessary to perform the time-consuming calculations in
equations (\ref{eq:SHint1}) and (\ref{eq:SHint2}) once (these might
be sped up by FFT-type fast algorithms which we did not use, see,
e.g. \cite{sph_fast}).
In order to find the rotation matrix
$\mathbf{R}$ that best aligns two distributions of $N$ particles
with respect to $L^2$, we may simply compute
\begin{equation}\label{eq:argminRot}
 \mathbf{R} = \underset{{\mathbf{R} \in SO(3)}}{\operatorname{argmin}}\,\, \mathbf{e}(f,g_\mathbf{R}).
\end{equation}
Our rotations of the spherical harmonics were performed using the
freely available software archive SHTOOLS available at
\url{www.ipgp.fr/~wieczor/SHTOOLS}, and we computed the best
$\mathbf{R}$ by exhaustively searching over $SO(3)$ with a mesh of
two degrees precision in each of the $\theta, \phi$ directions.
We thus obtained a good initial guesses, for each snapshot, of the
sought rotations, and subsequently used Newton iteration to more
accurately determine the optimal $\mathbf{R}$.

\subsection{Template-based alignment attempts}
\begin{figure}
\begin{center} \includegraphics[width=125mm]{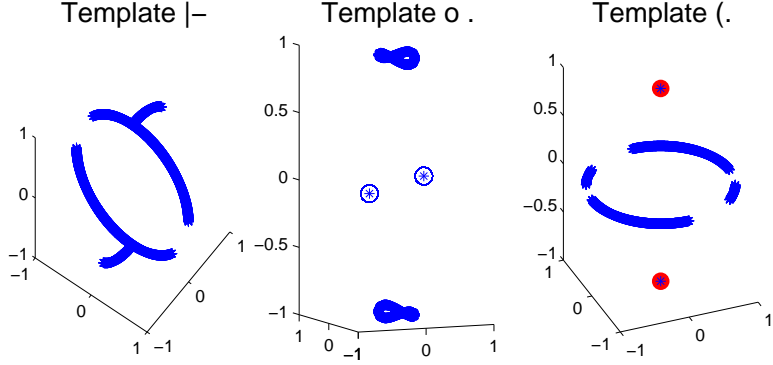} \end{center}
    \caption{The three templates utilized for alignment of the ensembles of $N$ particles.  Because Template $|-$ is vaguely reminiscent of the orientational distributions of Figures \ref{fig:init_dist_LCP} and \ref{fig:final_dist_LCP}, it is expected that it may do well (the ``matched filter'' is mathematically optimal, see \cite{matchFilter}).  Nevertheless, it drastically underperforms the eigenvector method, along with the other two templates (see Table \ref{table:LCP}).  Template o . consists of two ellipses (one centered at each pole of the unit sphere) with minor and major axes of $.2$ and $.4$, respectively, along with two points at $(\pm 1/ \sqrt{2}, \pm 1/ \sqrt{2},0)$, each with ``mass'' equal to one-quarter of the mass of the entire shape.  The red dots in Template ( . represent points with infinite weight and are located at $(0,0,\pm 1)$; the effect is that each snapshot is first rotated so that its center of mass lies along the z-axis, and then it is rotated along this axis (an $SO(2)$ rotation) to optimally align with the broken semicircle ``('' shape.}
    \label{fig:LCPtemplates}
\end{figure}

Using the spherical harmonics machinery, we first attempt to align
the set of $M$ ensembles of $N$ particles through the use
of fixed templates.
Somewhat arbitrarily, we chose the three fixed templates
shown in Figure \ref{fig:LCPtemplates}.
One of the template functions (Template $|-$) resembles the
orientational distributions of Figures \ref{fig:init_dist_LCP} and
\ref{fig:final_dist_LCP}; we anticipate that at least this template
will be useful in aligning the data.
Nevertheless, the global alignments
obtained with all three fixed templates fall short of those obtained with the
eigenvector method (see Table
\ref{table:LCP}).


\subsection{Application of the eigenvector method}
The first step in aligning the data through the eigenvector method
is to compute pairwise alignments (methodology discussed in Section
\ref{sec:SH}) between all $M$ distributions,
$\{\mathbf{R_{ij}}\}_{i,j=1}^M$.
Here, $\mathbf{R}_{ij}$ is the $3 \times 3$ matrix which rotates
ensemble $j$ to ensemble $i$.
Next, these pairwise rotations are
inserted in a large $3M \times 3M$ matrix of the following form:
\begin{equation}
 \mathbf{M} = \left[\begin{array}{cccc} \mathbf{R}_{11} & \mathbf{R}_{12} & \cdots & \mathbf{R}_{1M} \\
\mathbf{R}_{21} & \mathbf{R}_{22} & \cdots & \mathbf{R}_{2M} \\
\vdots & \vdots & \ddots & \vdots \\
\mathbf{R}_{M1} & \mathbf{R}_{M2} & \cdots & \mathbf{R}_{MM} \end{array} \right].
\end{equation}
In an ideal setting with no misalignments and no noise, the $ij$-th
block of the matrix $\mathbf{M}$ would simply be $\mathbf{R}_i
\mathbf{R}_j^T$, for this is the matrix which takes distribution $j$
back to the standard axes, and then rotates it by $\mathbf{R}_i$ in
order for it to coincide with distribution $i$.
We also note that in this ideal setting, the following equation
holds:
\begin{equation}\label{eq:rotEV}
 \mathbf{M} \mathbf{v} = M \mathbf{v},
\end{equation}
where $\mathbf{v}$ is the $3M \times 3$ matrix
\begin{equation}
 \mathbf{v} = \left[\begin{array}{c} \mathbf{R}_1 \\ \mathbf{R}_2 \\ \vdots \\ \mathbf{R}_M \end{array} \right].
\end{equation}
Therefore, the top three eigenvectors of $\mathbf{M}$ (each with
eigenvalue $M$) contain information about the ``unknown'' rotation
matrices $\mathbf{R}_i$.  Here, the matrix $\mathbf{M}$ is of rank
$3$, and $\mathbf{M} = \mathbf{v} \mathbf{v}^T$. That is,
$\mathbf{M}$ has only two distinct eigenvalues: an eigenvalue of $M$
whose multiplicity is $3$, and an eigenvalue of $0$ whose
multiplicity is $3M-3$. It is therefore expected that the top three
eigenvectors would not be affected too much by noise and
misalignments.

When there is some noise, the $\mathbf{v}$ matrix will still be
resolved (but now with eigenvalues slightly less than $M$), and we
expect to be able to recover the information contained in these
columns regarding the rotation matrices $\mathbf{R}_i$.
The recovery will
be, of course, only up to an orthogonal transformation which is an
inherent degree of freedom:  by specifying only pairwise rotations,
one only knows how the distributions look relative to each other.
This transformation (in effect, its three associated degrees of
freedom) appears in equation (\ref{eq:rotEV}), for this equation
holds not only for $\mathbf{v}$, but also for $\mathbf{v}
\mathbf{R}$ for any $\mathbf{R} \in \mbox{SO}(3)$.
Due to the noise, each recovered $\mathbf{R}_i$ is not exactly a
rotation matrix (this phenomenon is analogous to the $z_i$ not being
necessarily of unit length in equation (\ref{eq:maxZ}) of Section
\ref{sec:SummaryMethod}).
However, one can find the closest (in Frobenius norm) rotation
matrix via the well-known procedure: $\mathbf{R}_i \rightarrow
\mathbf{U}_i \mathbf{V}_i^T$, where $\mathbf{U}_i \mathbf{\Sigma}_i
\mathbf{V}_i^T$ is the SVD of $\mathbf{R}_i$
(\cite{Fan1955,Keller1975}).

\begin{figure}
\begin{center} \includegraphics[width=125mm]{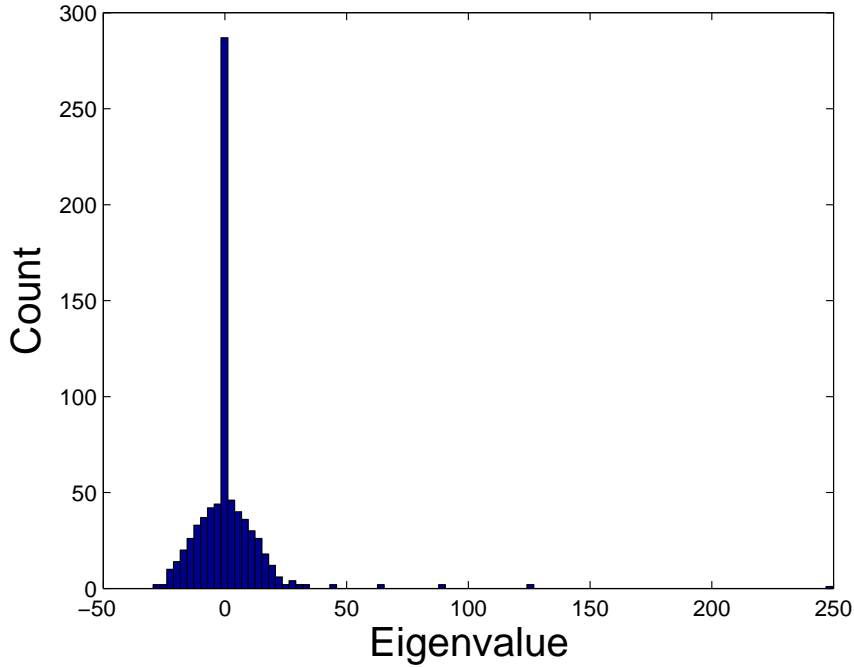} \end{center}
    \caption{A histogram for the eigenvalues for $M=250$ and $N=10$.  One eigenvalue is near $250$, and there are four pairs of eigenvalues near $125$, $88$, $63$, and $44$ (these are likely due to the correlations in the matrix $\mathbf{M}$, see \cite{Xiuyuan} and Section \ref{sec:SummaryMethod}).  The other eigenvalues are ``in the semicircle.''  We need the largest three eigenvalues and their eigenvectors to recover the rotations.  So, in practice we use the eigenvalue near $250$ and the pair of eigenvalues near $125$.}
    \label{fig:SHeigs}
\end{figure}

Because \cite{feral} requires the noise in each element of the matrix $\mathbf{M}$
to be independent, we do not see the expected semicircle-type  distribution in
our Figure \ref{fig:SHeigs}; notice, however, that a shape reminiscent of a semicircle can
still be seen.
Again, this is due to the fact that ``good'' and ``bad''
measurements are not random and independent, but rather, correlated;
having independent entries requires $M^2$ sources of randomness, and
clearly, for large $M$, this is not true because the ``amount'' of
randomness grows only as $M$, the number of snapshots.  Because of
this, for large $M$, some eigenvalues of $O(M)$ appear outside the
semicircle.  See \cite{Xiuyuan} for details.

%
The error in the global rotations recovered are shown in Table
\ref{table:LCP}.
The eigenvector method appears quite successful: even for large
amounts of noise (small $N$) and small values of $M$, favorable
results are obtained.
Even though the eigenvalue semicircle analysis of Section
\ref{sec:SummaryMethod} was carried out for group $SO(2)$ and not
the group $SO(3)$ of interest here, the distance from the leading
eigenvector to the noisy semicircle still quantifies the alignment
error.
Furthermore, as expected, when both $M$ and $N$ both become large
(large $N$ means that the probability of ``good'' measurements goes
up, and in the context of Section \ref{sec:SummaryMethod}, that $p
\rightarrow 1$), the leading eigenvalues ($\lambda_{H1},
\lambda_{H2}, \lambda_{H3}$) increase as $O(M)$ and the position of
$\lambda_R$ increases as $O(\sqrt{M})$.
These results are summarized in Table \ref{table:LCP}.

\begin{table}
\begin{center}
\begin{tabular}{ c | c | c | c | c | c | c | c | c | c} 
  $N$ & $M$ & $\lambda_{H1}$ & $\lambda_{H2}$ & $\lambda_{H3}$ & $\lambda_{R}$ & $\mbox{err}_{\lambda}$ & $\mbox{err}_{|-}$ & $\mbox{err}_{o .}$ & $\mbox{err}_{( .}$ \\ \hline
       & $250$ & $249.9$ & $182.9$ & $182.9$ & $22.2$ & $0.078$  & & & \\
       & $125$ & $125.0$ & $91.2$  & $91.2$  & $18.1$ & $0.125$  & & & \\
$5000$ & $63$  & $63.0$  & $47.8$  & $47.8$  & $11.7$ & $0.189$  & $1.3671$ & $2.0801$ & $1.7671$ \\
       & $32$  & $32.0$  & $23.6$  & $23.5$  & $7.9$  & $0.196$  & & & \\
       & $16$  & $16.0$  & $13.0$  & $13.0$  & $4.6$  & $0.204$  & & & \\ \hline

       & $250$ & $249.9$ & $179.2$ & $179.2$ & $24.3$ & $0.110$  & & & \\
       & $125$ & $125.0$ & $89.9$  & $89.9$  & $19.6$ & $0.135$  & & & \\
$1000$ & $63$  & $63.0$  & $44.8$  & $44.7$  & $14.4$ & $0.197$  & $1.4339$ & $2.2302$ & $1.9340$ \\
       & $32$  & $32.0$  & $23.4$  & $23.3$  & $8.6$  & $0.225$  & & & \\
       & $16$  & $16.0$  & $12.8$  & $12.8$  & $5.9$  & $0.351$  & & & \\ \hline

       & $250$ & $249.9$ & $175.9$ & $175.9$ & $25.8$ & $0.181$  & & & \\
       & $125$ & $125.0$ & $88.2$  & $88.1$  & $19.7$ & $0.193$  & & & \\
$200$  & $63$  & $63.0$  & $44.4$  & $44.4$  & $14.6$ & $0.257$  & $1.4398$ & $2.4334$ & $1.9420$ \\
       & $32$  & $32.0$  & $23.3$  & $23.3$  & $9.4$  & $0.303$  & & & \\
       & $16$  & $16.0$  & $12.8$  & $12.8$  & $6.5$  & $0.358$  & & & \\ \hline

       & $250$ & $249.9$ & $146.2$ & $146.2$ & $36.2$ & $0.403$  & & & \\
       & $125$ & $124.9$ & $75.6$  & $75.6$  & $19.9$ & $0.412$  & & & \\
$50$   & $63$  & $63.0$  & $39.0$  & $39.0$  & $15.3$ & $0.535$  & $1.4419$ & $2.5908$ & $2.0040$ \\
       & $32$  & $32.0$  & $21.2$  & $21.1$  & $10.2$ & $0.624$  & & & \\
       & $16$  & $16.0$  & $9.9$   & $9.9$   & $9.9$  & $0.780$  & & & \\ \hline

       & $250$ & $248.8$ & $125.1$ & $125.1$ & $37.2$ & $0.604$  & & & \\
       & $125$ & $124.9$ & $63.2$  & $63.2$  & $21.5$ & $0.836$  & & & \\
$20$   & $63$  & $63.0$  & $31.7$  & $31.7$  & $19.3$  & $0.862$ & $1.4475$ & $2.6133$ & $2.0084$ \\
       & $32$  & $32.0$  & $17.8$  & $17.8$  & $10.3$  & $0.988$ & & & \\
       & $16$  & $16.0$  & $9.6$   & $9.6$   & $10.2$  & $1.087$ & & & \\ \hline

       & $250$ & $249.7$ & $124.2$ & $124.1$ & $38.8$ & $1.148$  & & & \\
       & $125$ & $124.8$ & $61.8$  & $61.7$  & $22.4$ & $1.192$  & & & \\
$10$   & $63$  & $62.9$  & $30.0$  & $30.0$  & $22.0$ & $1.195$  & $1.5413$ & $2.6156$ & $2.3441$ \\
       & $32$  & $31.9$  & $16.9$  & $16.9$  & $11.1$ & $1.263$  & & & \\
       & $16$  & $16.0$  & $8.9$   & $8.9$   & $10.9$ & $1.459$  & & & \\

\end{tabular}
\caption{Results of the eigenvector method (vs. $N$ and $M$) and
alignments with various templates (vs. $N$).  As before, $N$ is the
number of LCP particles representing the distribution (smaller $N$
implies more noise).  $M$ is the number of ensembles of $N$
particles, meaning that we perform $M(M-1)/2$ comparisons when using
the eigenvector method.  $\lambda_{H1}$, $\lambda_{H2}$, and
$\lambda_{H3}$ are the $3$ largest eigenvalues of the $3M \times 3M$
matrix $\mathbf{M}$, and these are the eigenvalues which contain the
rotation matrix information.  $\lambda_R$ is the eigenvalue at the
right edge of the semicircle.  Finally, the ``err'' quantities
describe the error in the computed $\mathbf{R}_i$, and are equal to
$1/M \sum_{i=1}^M \| \mathbf{R}_{i \, true} - \mathbf{R}_i \|_F^2$
($\mbox{err}_\lambda$ is for the eigenvector method, and the rest
are for the templates shown in Figure \ref{fig:LCPtemplates}). The
eigenvector method easily outperforms all templates, and as
expected, the error grows as $M$ and $N$ get smaller due to less
information and more noise, respectively.  $\lambda_{H1}$ appears to
scale as $M$, while $\lambda_{H2}$ and $\lambda_{H3}$ scale with $M$
\textit{and} decrease with noise.  $\lambda_R$ increases with both
noise and $M$ (see Section \ref{sec:SummaryMethod}).
\label{table:LCP}}
\end{center}
\end{table}

\section{The second illustrative example: modulated traveling waves of the one-dimensional Kuramoto-Sivashinsky
equation} \label{sec:KSE}

Symmetries play an important role in systems that exhibit spatiotemporal pattern formation
(and the evolution equations that model them).
When processing experimental or computational data that arise in
observing such problems, it again makes sense to first factor out
the underlying symmetries.
As an example of such a spatiotemporal pattern-forming system, we choose the
Kuramoto-Sivashinksy equation (KSE) in one spatial dimension and with
periodic boundary conditions, which can be written in the following form:
\begin{eqnarray}
 u_t + 4 u_{xxxx} + \alpha [u_{xx}+uu_x] &=& 0, \notag \\
u(t,x) &=& u(t,x+2\pi). \label{eq:KSE}
\end{eqnarray}
This well-known nonlinear PDE arises as a model in many physical
contexts, from flame front propagation to the dynamics of falling
liquid films (\cite{ksflame,ksfilm}).
It gives rise to a rich variety of spatiotemporal dynamical patterns
including steady state multiplicity and symmetry-breaking
bifurcations, as well as traveling, modulated and ``turbulent"
waves.
It has been shown, under certain conditions, to possess inertial
manifolds (\cite{aim}), implying that its long-term dynamics are
low-dimensional; this low dimensionality, along with the rich
spatiotemporal dynamics, is an important reason for selecting it as
an illustrative example.

\subsection{System setup}

For certain values of the parameter $\alpha$, the KSE exhibits
attractors that are traveling waves that are not of constant shape,
but rather exhibit spatiotemporal fluctuations; these are termed
Modulated Traveling Waves (MTWs).
Such attractors can be thought of as two-dimensional tori ($T^2$)
in infinite-dimensional space; one ``direction" around the torus
corresponds to traveling, and the other to a periodic modulation.
We will study {\em transient} computational data obtained in such
a parameter regime; the data do not necessarily lie {\em on} the
MTW attractors, but they are visually close enough that the two types
of motion are visible in our plots.

Equation (\ref{eq:KSE}) is equivariant with respect to spatial
translations; therefore, the ``traveling'' behavior of these waves
may be factored out (the underlying symmetry group is that of
positions $x$ modulo $2 \pi$, or, as we referred to it before, that
of angles modulo $2 \pi$ -- $SO(2)$).
Writing equation (\ref{eq:KSE}) as $u_t = \mathcal{D}(u)$, the
equivariance relation becomes $\mathcal{D}(S_c[v]) =
S_c[\mathcal{D}(v)]$, where $S_c[v](x) = v(x+c)$ is the
\textit{shift operator} on spatially periodic functions.

Although the traveling behavior of the wavy transients can be
factored out, their modulation \textit{cannot}.
For an exact MTW attractor, where {\em the modulation} (as opposed
to the traveling) is exactly periodic in time, there does exist a
continuous, one-to-one map between each phase of the
\textit{temporal} modulation and the set of points on the circle;
yet this does not lead to equivariance.
It is the \textit{spatial} shifts of arbitrary wave profiles (not
the temporal ones on {\em exactly} periodic attractors) that we are
interested in.

\begin{figure}
\begin{center} \includegraphics[width=125mm]{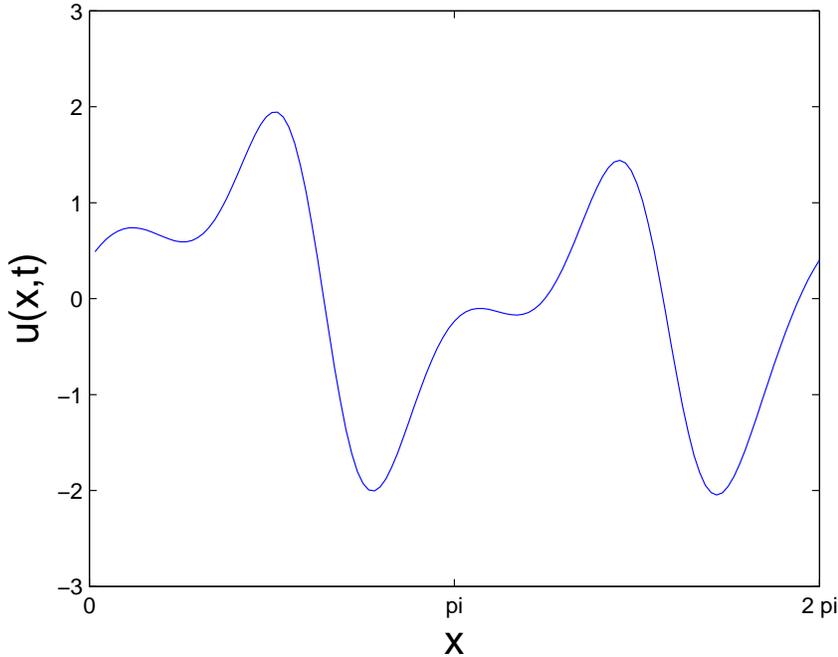} \end{center}
    \caption{A temporal snapshot (spatial profile at a moment in time) from a PDE
solution close to a modulated traveling wave attractor for $\alpha =
32$.}
    \label{fig:MTW}
\end{figure}

\begin{figure}
\begin{center} \includegraphics[width=125mm]{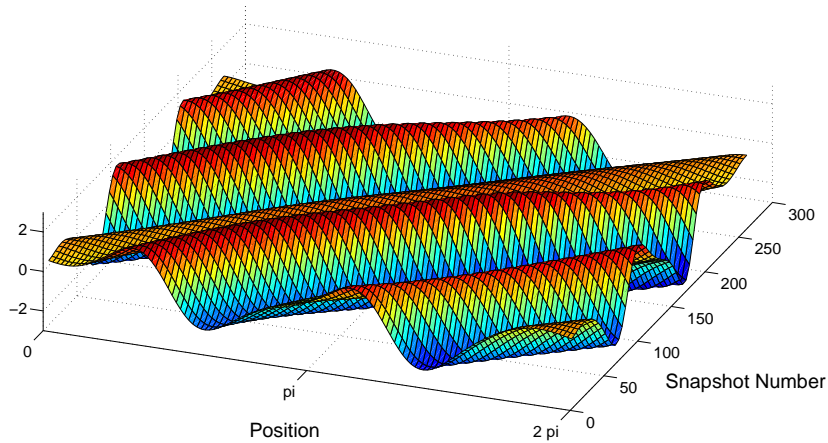} \end{center}
    \caption{A sequence of MTW snapshots demonstrating the temporal evolution of the MTW.  One can clearly see the
traveling motion and a slight modulation on top of this motion.}
    \label{fig:MTWseq}
\end{figure}

An additional qualitative computational observation is that variations in the
solution snapshots associated with the traveling component of the
evolution are
significantly larger than the variations associated with the ``modulation" part,
which remains after the traveling has been factored out (as will be
described below).
Based on this observation, we will still use the eigenvector method
to align the data, even though in its formulation such a modulation
is not taken into account (for a formulation which {\em does} take
this into account, see the discussion about vector diffusion maps,
Section \ref{sec:vDMAPS}, further below).
We will compare this to alignments obtained using template-based methods
(as was done in \cite{rowley2}).
The output of both methods, the list of global alignments for each
wave snapshot, can then be used to align each wave snapshot so that
the traveling motion is factored out and we can focus on studying
the modulation exclusively (for instance, through the use of
diffusion maps).
Figure \ref{fig:MTW} is a picture of (a transient closely approximating) a modulated traveling wave, and
Figure \ref{fig:MTWseq} shows the temporal evolution of the wave shapes on this transient.

\subsection{Generation of snapshot data in the MTW parameter regime}

To generate an ensemble of $M$ transient snapshots in the neighborhood of
an MTW attractor, we begin by
integrating equation (\ref{eq:KSE}) for an extended period of time
on an evenly spaced grid of $128$ mesh points with width $2 \pi /
128$.
Because the MTW behavior is an attractor for the
system, after this long time, each $u(x,t)$ at a fixed
time $t$ can be thought of as accurately approximating a snapshot {\em on} the MTW.
In Figure \ref{fig:MTWseq} we show a sequence of such ``MTW snapshots".

We take snapshots
$u(x,t_1),u(x,t_2),\hdots,u(x,t_M)$ at $M$ different times $t_1,t_2,\hdots,t_M$.
We then make these snapshots artificially
noisy by adding Gaussian white noise of variance $\sigma^2$ to each
of them (to each of the $128$ mesh points $\{ x_i \}_{i=1}^{128}$,
we add a normal random variable of variance $\sigma^2$).
Without this noise, traditional single template-based approaches can
do a very good job of factoring out the traveling motion of the MTW.
With this noise, however,
template-based approaches can fail spectacularly, while the eigenvector alignment method
may still usefully resolve the global alignments.

\subsection{Alignment of MTW snapshots}

To find the alignment $a_{ij}$ which aligns a (noisy) wave snapshot
$u(x,t_i)$ with another $u(x,t_j)$, we simply set $a_{ij}$ equal to
the $k \in  \{1,2,\hdots,128 \}$ which minimizes
\begin{equation} \label{eq:maxCorr}
 \sum_{l=1}^{128} \left[ T_k [u](x_l,t_i)-u(x_l,t_j) \right]^2;
\end{equation}
here $T_k$ is the periodic \textit{shift operator} on the $128$ mesh
points $\{x_i \}_{i=1}^128$ defined by $T_k [u](x_l,t) = u(x_{k+l},t)$.
The analogue of
equation (\ref{eq:maxCorr}) is also used to align a (noisy) MTW
snapshot against a chosen (fixed) template.

\subsection{Template-based alignment attempts}

%
Although it is
best to select a template with some prior knowledge, even relatively arbitrary
choices (e.g. a ``Mexican hat") may give good results.
When the wave snapshots contain more
than a little noise, however, alignment with a template will certainly
give rise to many incorrect answers.
Furthermore, even with no noise, poor template
choices may result in spurious alignments.

\begin{figure}
\begin{center} \includegraphics[width=125mm]{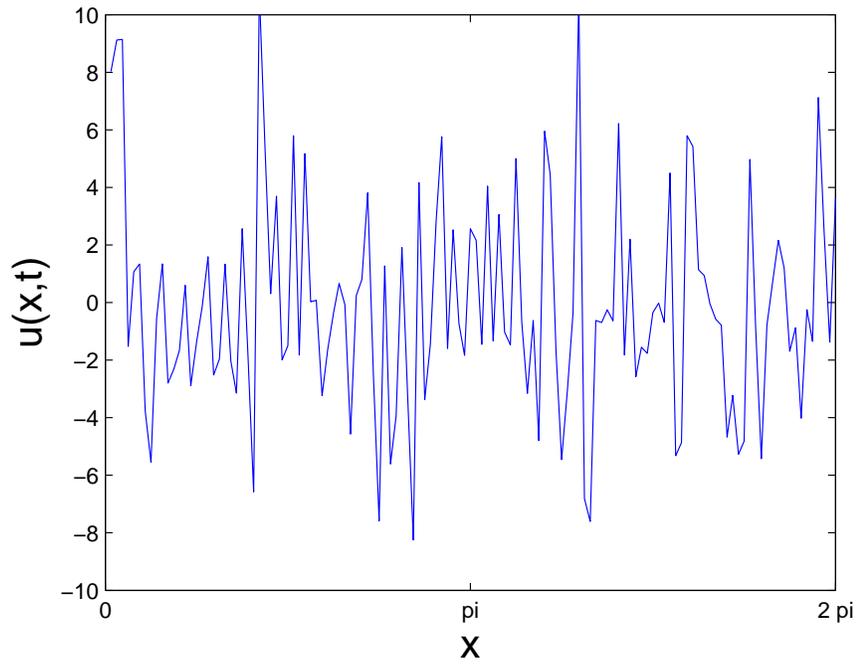} \end{center}
    \caption{A snapshot of an MTW with Gaussian white noise of variance $3.5^2$ added.}
    \label{fig:MTWnoise}
\end{figure}

Figure \ref{fig:MTWnoise} shows a MTW snapshot with added Gaussian white
noise of variance $\sigma^2 = 3.5^2$.
Clearly, this amount of noise will present a problem to alignment efforts:
it is difficult to even visually perceive the resemblance with the
noiseless MTW snapshot (Figure \ref{fig:MTW}).
Nevertheless, we
attempt to align this snapshot (as well as others taken from our data
set, with the same type of noise added) using different templates.
The next
series of figures shows
\begin{itemize}
 \item on the right, the alignment of each noisy wave snapshot in our data set with a particular single template
 (obtained by finding the periodic shift which, according to equation (\ref{eq:maxCorr}), results in maximum correlation/minimum $L^2$ distance with the fixed template) vs. its correct alignment; ideally, this plot should consist of one straight line (after taking into account periodicity)
 \item on the left, to demonstrate the degree of robustness of the alignment procedure, a plot of the $L^2$ distance
 between the template function and ``all"  periodic shifts of \textit{a single} noisy MTW snapshot randomly chosen from our data set. This
 function's minimum is the alignment chosen for this noisy MTW by the template method (it maximizes the correlation/minimizes the $L^2$ distance),
 and it is this ``best" alignment for all the snapshots that is plotted in the figure on the right.
\end{itemize}

\begin{figure}
\begin{center} \includegraphics[width=125mm]{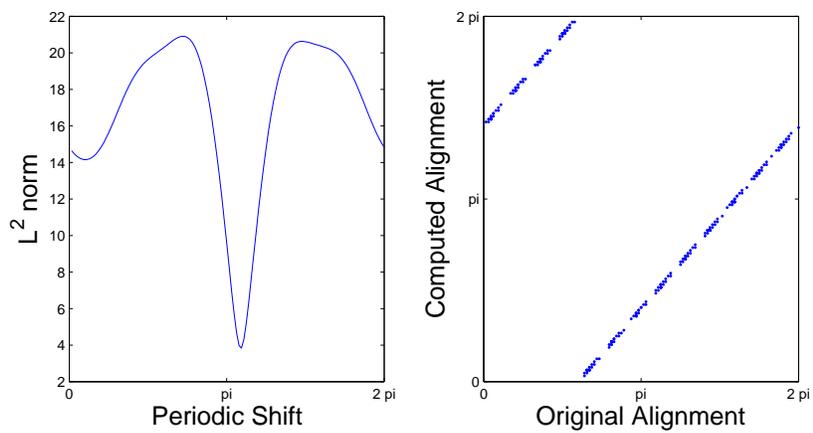} \end{center}
    \caption{Alignment of noiseless MTW snapshots with a single noiseless MTW as the template.  As expected, the alignment appears perfect.}
    \label{fig:nonoise2align}
\end{figure}

\begin{figure}
\begin{center} \includegraphics[width=125mm]{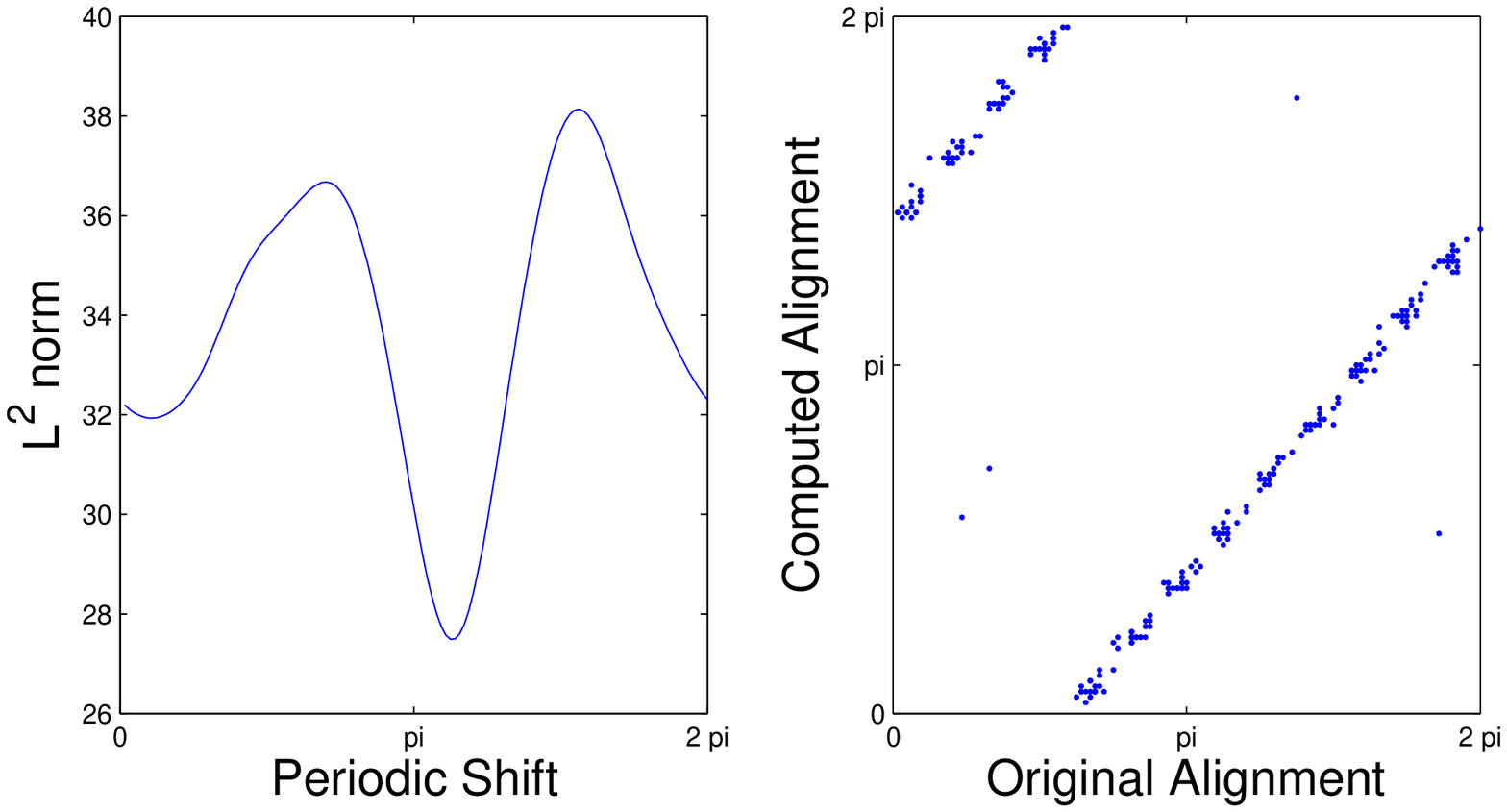} \end{center}
    \caption{Alignment of noisy MTW snapshots with a single noiseless MTW as the template.  The resulting alignment is nearly perfect, but one can see that the robustness of Figure \ref{fig:nonoise2align} has already started to wane; the range of the $L^2$ distances computed in this figure is much narrower than that of Figure \ref{fig:nonoise2align}.}
    \label{fig:nonoisealign}
\end{figure}

First, for reference, the alignment of \textit{noiseless} snapshots
with a template (here, the template was chosen to be a particular
noiseless MTW snapshot) would appear like Figure
\ref{fig:nonoise2align}.
In this figure, as expected, the alignments obtained are nearly
perfect (the figure resembles a straight line with small-in the
$L^2$ norm-``gaps'' caused by the modulation, which we will not
study further here).
These favorable results are expected since we are using a
mathematically motivated template (a ``matched filter,'' see
\cite{matchFilter}) in noiseless conditions.
In a slightly more
realistic setting our snapshots will be noisy (and we still use a noiseless
MTW snapshot as our template);
this result is shown in Figure
\ref{fig:nonoisealign}.
Again, the alignments obtained are nearly perfect (the  small
``gaps'' also remain), but now there are a few errors.
Of course, using a noiseless MTW snapshot as our template can be
thought of as slightly ``cheating''; from our data set of noisy
waves close to an MTW attractor, we do not know what an exact,
noiseless MTW snapshot looks like.

We tried several other template functions, including the Mexican
hat, a cosine function, a step function (equivalently, the second
Haar wavelet), and a triangle.
Voting-based approaches were also tried;
in these approaches, the results of multiple templates were averaged
together in a suitable way in order to come up with a consensus.
These voting-based approaches were also seen to fail; knowing how to average
the votes together is a problem, and some templates have many local
minima.
Finally, center of mass- and moment-based alignment approaches also
appeared to fail; this was not unexpected, since aligning based on
moments is closely related to template alignmnent.
Some of these figures are shown below.

\begin{figure}
\begin{center} \includegraphics[width=125mm]{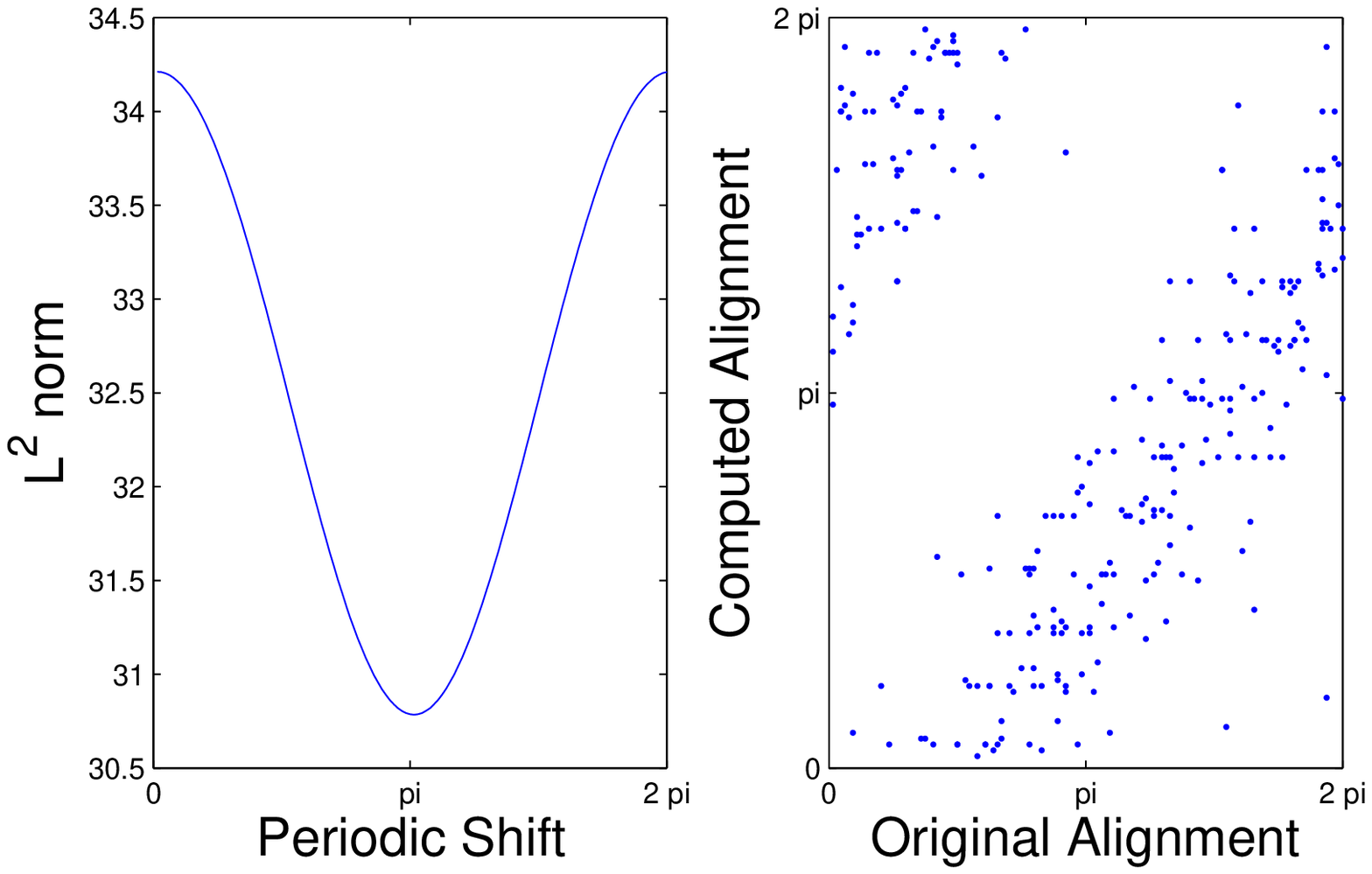} \end{center} \label{fig:MTWcos}
    \caption{Alignment of noisy MTW snapshots with a cosine.  Although the $L^2$ distance plot (left) is smooth, its range is \textit{much} narrower than that of Figure \ref{fig:nonoise2align} leading to poorer alignments.}
\end{figure}
%
%
\begin{figure}
\begin{center} \includegraphics[width=125mm]{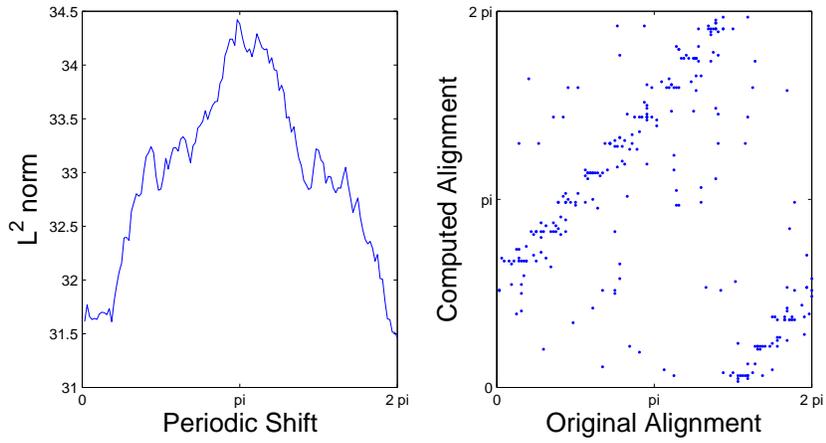} \end{center} \label{fig:MTW_2ndHaar}
    \caption{Alignment of noisy MTW snapshots with a step function (the second Haar wavelet).  Here, the $L^2$ distance plot exhibits a narrow range of values.}
\end{figure}
%
%
\begin{figure}
\begin{center} \includegraphics[width=125mm]{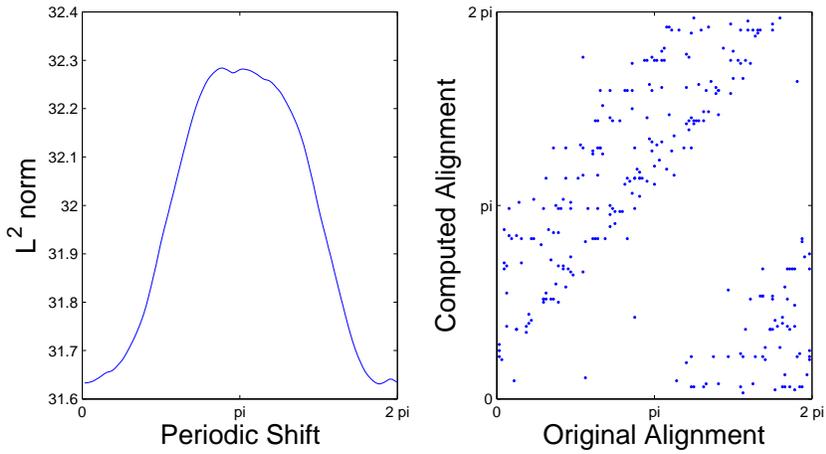} \end{center}
   \caption{Alignment of noisy MTW snapshots with a triangle-shaped template.  As in Figures \ref{fig:MTWcos} and \ref{fig:MTW_2ndHaar}, the $L^2$ distance range is narrow and the alignments obtained are poor.}
\end{figure}


The only template to give a visually satisfactory answer was a (in
principle, unavailable) noiseless MTW snapshot (again, see Figure
\ref{fig:nonoisealign}).
Since the noiseless MTW template gave such good results, one might
be tempted to try a \textit{noisy} MTW from the data set as a
template (which would \textit{not} be considered cheating!).
However, the performance of such a template is
spectacularly poor: see Table \ref{table:KSE} for summary
statistics.


\subsection{Application of the eigenvector method}
\label{sec:KSEeVec}

In the presence of so much noise (again, see Figure
\ref{fig:MTWnoise}), it is difficult to imagine aligning the noisy
wave snapshots without prior knowledge of a good template such as
the one provided by a noiseless MTW snapshot (Figure
\ref{fig:nonoisealign}).
However, the eigenvector method takes into account information based
on all pairwise alignments (in essence, treating \textit{each} wave
snapshot as a template, and looking at all $M(M-1)/2$ comparisons)
and it is seen to give surprisingly good results.

First, we compute all pairwise alignments between the $M$ noisy wave snapshots
by finding the alignment $a_{ij}$ which
minimizes the corresponding $L_2$ norm of their difference; clearly,
many of these may be computed incorrectly.
The alignment which rotates snapshot $i$ to snapshot $j$, denoted
$a_{ij}$, is (for our spatially discretized waveforms) an integer
between $1$ and $128$, describing how many mesh points forward one
must shift snapshot $i$ in order for it to maximally correlate with
snapshot $j$.
This alignment is then mapped to the
unit circle via $\mathbf{T}_{ij} = \mbox{exp}(-2i \pi a_{ij}/128)$,
and a matrix $\mathbf{T}$ is constructed as follows:
\begin{equation}
\mathbf{T} = \left[\begin{array}{cccc} \mbox{exp}(-2i \pi a_{11}/128)&\mbox{exp}(-2i \pi a_{12}/128)&\cdots&\mbox{exp}(-2i \pi a_{1M}/128) \\
\mbox{exp}(-2i \pi a_{21}/128)&\mbox{exp}(-2i \pi a_{22}/128)&\cdots&\mbox{exp}(-2i \pi a_{2M}/128)\\
\vdots & \vdots & \ddots & \vdots \\
\mbox{exp}(-2i \pi a_{M1}/128)&\mbox{exp}(-2i \pi a_{M2}/128)&
\cdots & \mbox{exp}(-2i \pi a_{MM}/128) \end{array} \right].
\end{equation}
In an ideal setting with no noise/no misalignments, the $ij$-th
block of the matrix $\mathbf{T}$ would simply be $\mbox{exp}[-2i
\pi/128 (a_j-a_i)]$ (where we denote the actual, unknown rotation of
snapshot $i$ by $a_i$); this is the rotation which takes snapshot
$j$ back to the ``phase" zero, and then rotates it by $\mbox{exp}(2i
\pi a_i / 128)$ in order for it to coincide with snapshot $i$.
We also note that, in this ideal setting, the following equation
holds:
\begin{equation} \label{eq:rotEV2}
 \mathbf{T} \mathbf{v} = M \mathbf{v},
\end{equation}
with
\begin{equation}
 \mathbf{v} = \left[\begin{array}{c} \mbox{exp}(2i\pi a_1/128) \\ \mbox{exp}(2i\pi a_2/128) \\ \vdots \\ \mbox{exp}(2i\pi a_M/128) \end{array} \right].
\end{equation}

The top eigenvector of $\mathbf{T}$ (with eigenvalue $M$) contains,
therefore, information about the shifts $a_i$ (the ``alignments").
In this setting, the matrix $\mathbf{T}$ is of rank $1$, and it
satisfies $\mathbf{T} = \mathbf{v} \mathbf{v}^T$, so $\mathbf{T}$
has two distinct eigenvalues:  an eigenvalue of $M$ whose
multiplicity is $1$ and an eigenvalue of $0$ whose multiplicity is
$M-1$.  It is therefore expected that the top eigenvalue would not
be affected too much by noise and misalignments.

When there is some noise, $\mathbf{v}$ will still be approximately
resolved (but now with eigenvalue slightly less than $M$), and we
are able to recover the information contained in this eigenvector
regarding the alignments $a_i$.
The recovery will be, of course, only up to an overall global
shift, which (since we only specify pairwise
relative shifts) is an inherent degree of freedom.
This can be seen in equation (\ref{eq:rotEV2}); this
equation holds for not only $\mathbf{v}$, but also for $\mbox{exp}(i
\theta) \mathbf{v}$ (and, in fact, any constant times $v$).
In fact,
due to the noise, each recovered $a_i$ will not have exactly unit magnitude;
yet the $a_i$ may be recovered by considering both the imaginary and
real parts of the $i$th entry of $\mathbf{v}$.
In particular, we
set
\begin{equation}
a_i = \frac{128}{2 \pi} \times \mbox{arctan}
\left(\frac{\mbox{Im}(\mathbf{v}_i)}{\mbox{Re}(\mathbf{v}_i)}\right).
\end{equation}



\begin{figure}
\begin{center} \includegraphics[width=125mm]{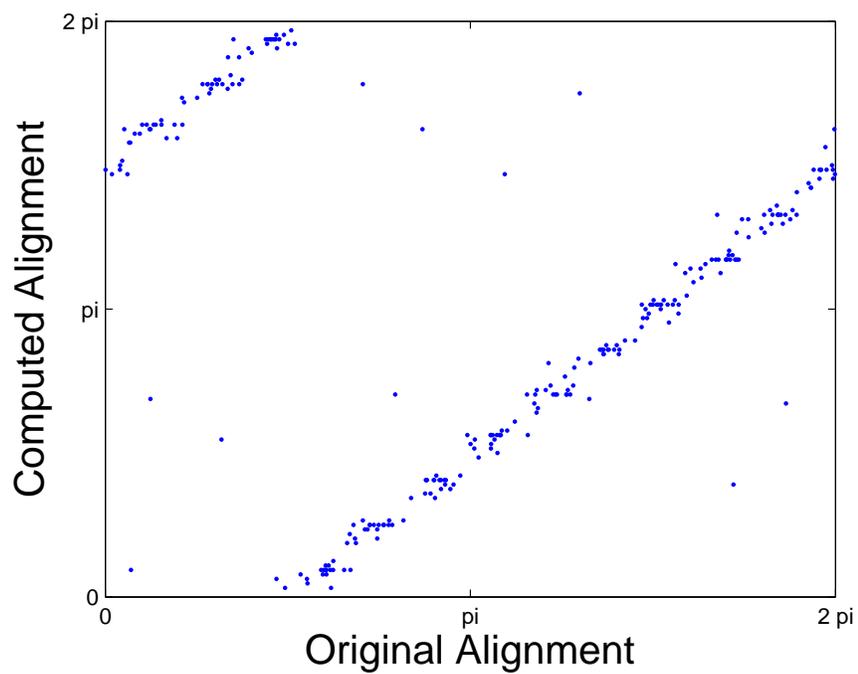} \end{center}
    \caption{Noisy MTW snapshot alignments obtained using the eigenvector method.  A significant improvement upon single template methods is observed.}
    \label{fig:synch}
\end{figure}

\begin{figure}
\begin{center} \includegraphics[width=125mm]{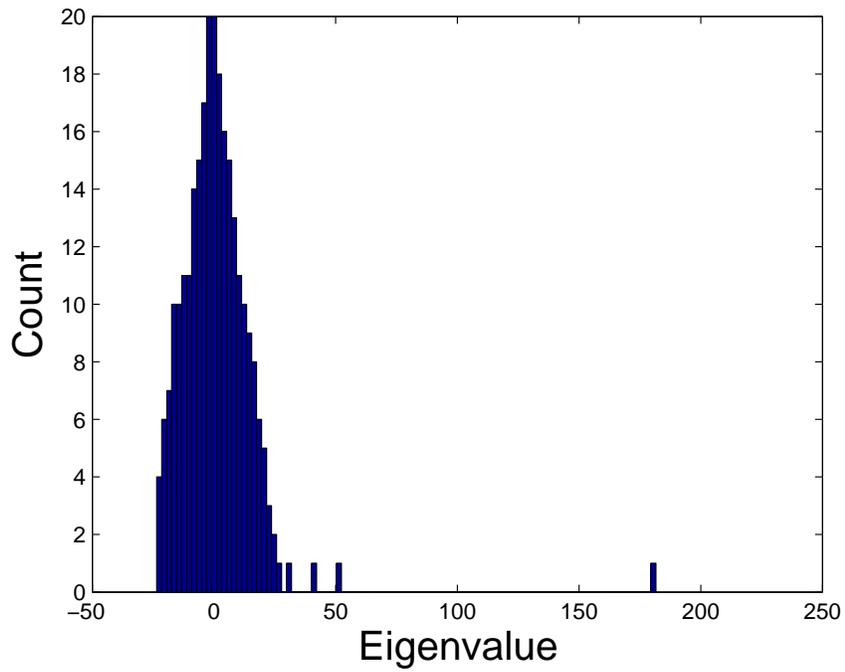} \end{center}
    \caption{A histogram for the eigenvalues for $M=250$ and $\sigma^2 = 2.5^2$. One dominant eigenvalue is near $180$, and there are two others near $50$ (these are likely due to the correlations in the matrix $\mathbf{M}$, see \cite{Xiuyuan} and Section \ref{sec:SummaryMethod}).  The rest of the eigenvalues appear to belong to the ``semicircle.''}
    \label{fig:eval_KSE}
\end{figure}

The results of the eigenvector method constitute, without a doubt, a
significant improvement upon those obtained using the various fixed
templates above (see Figure \ref{fig:synch} and Table
\ref{table:KSE}.
The eigenvalue
distribution can be seen in Figure \ref{fig:eval_KSE}; one large
eigenvalue clearly dominates the rest.
However, because the theory in \cite{feral} requires the noise in
each of the elements of the  matrix $\mathbf{M}$ to be independent,
we do not see the predicted semicircle distribution in Figure
\ref{fig:eval_KSE} (although a shape reminiscent of the semicircle
can still be discerned).
Again, this is due to the fact that ``good'' and ``bad''
measurements are not random, but rather, correlated; having
independent entries requires $M^2$ sources of randomness, and
clearly, for large $M$, this is not true because there are only $128
\times M$ sources ($M$ snapshots and $128$ random Gaussian variables
for each snapshot).
Therefore, for large $M$, some eigenvalues of magnitude $O(M)$
appear {\em outside} the ``semicircle"; see \cite{Xiuyuan} for
details.

For even larger amounts of noise and even smaller values of $M$,
good results can still be obtained.
In fact, the distance from the leading eigenvalue to the ``noisy
semicircle" quantifies the alignment error (see Section
\ref{sec:SummaryMethod}).
When $M$ is large and the problem is relatively noiseless (so that
in the context of Section \ref{sec:SummaryMethod}, $p \approx 1$),
the distance from $\lambda_H$ to $\lambda_R$ is predicted to be
large (again, see Section \ref{sec:SummaryMethod}); the position of
the leading eigenvalue $\lambda_H$ increases as $O(M)$ and the
position of $\lambda_R$ increases as $O(\sqrt{M})$.

\begin{table}
\begin{center}
\begin{tabular}{ c | c | c | c | c | c | c | c | c | c | c | c | c }
  $\sigma^2$ & $M$ & $\lambda_H$ & $\lambda_R$ & $\rho_{\lambda}'$ &
  $\rho_{\lambda}$ & $\rho_{cos}$ & $\rho_{2m}$ & $\rho_{tr}$ & $\rho_{mh}$ & $\rho_{2H}$ & $\rho_{M}$ & $\rho_{nM}$
  \\ \hline
        & $250$ & $248.1$ & $9.9$  & $1.00$ & $1.00$  & & & & & & & \\
        & $125$ & $123.6$ & $4.1$  & $1.00$ & $.999$  & & & & & & & \\
$1.0^2$ & $62$  & $60.6$  & $2.6$  & $.999$ & $.999$  & $.930$ & $.837$ & $.779$ & $.340$ & $.992$ & $.999$ & $.996$  \\
        & $32$  & $31.7$  & $1.3$  & $.999$ & $.999$  & & & & & & & \\
        & $16$  & $15.9$  & $0.6$  & $.999$ & $.999$  & & & & & & & \\ \hline
        & $250$ & $217.9$ & $24.5$ & $.994$ & $.978$  & & & & & & & \\
        & $125$ & $110.2$ & $14.3$ & $.993$ & $.976$  & & & & & & & \\
$2.0^2$ & $62$  & $55.1$  & $10.1$ & $.993$ & $.972$  & $.798$ & $.556$ & $.663$ & $.257$ & $.846$ & $.994$ & $.918$   \\
        & $32$  & $28.3$  & $5.4$  & $.988$ & $.970$  & & & & & & & \\
        & $16$  & $14.0$  & $2.3$  & $.979$ & $.967$  & & & & & & & \\ \hline
        & $250$ & $181.9$ & $25.1$ & $.974$ & $.956$  & & & & & & & \\
        & $125$ & $93.9$  & $19.4$ & $.974$ & $.953$  & & & & & & & \\
$2.5^2$ & $62$  & $47.2$  & $13.8$ & $.965$ & $.952$  & $.757$ & $.639$ & $.673$ & $.272$ & $.783$ & $.982$ & $.882$  \\
        & $32$  & $24.4$  & $9.5$  & $.956$ & $.946$  & & & & & & & \\
        & $16$  & $13.2$  & $6.2$  & $.956$ & $.946$  & & & & & & & \\ \hline
        & $250$ & $124.9$ & $30.4$ & $.874$ & $.868$  & & & & & & & \\
        & $125$ & $61.2$  & $21.2$ & $.828$ & $.836$  & & & & & & & \\
$3.5^2$ & $62$  & $33.3$  & $14.8$ & $.794$ & $.832$  & $.602$ & $.436$ & $.504$ & $.254$ & $.625$ & $.909$ & $.514$  \\
        & $32$  & $18.0$  & $10.6$ & $.771$ & $.805$  & & & & & & & \\
        & $16$  & $9.86$  & $6.1$  & $.698$ & $.802$  & & & & & & & \\
\end{tabular}
\caption{Results of the eigenvector method (vs. $\sigma^2$ and $M$)
and the alignments with various templates (vs. $\sigma^2$). For the
eigenvector method, shown are the quantities $\lambda_H$ (the
largest eigenvalue), $\lambda_R$ (the right edge of the semicircle),
$\rho_{\lambda} = \left|\frac{1}{\sqrt{M}} \sum e^{-i \theta_i} v(i)
\right|$, and $\rho_{\lambda}' = \left|\frac{1}{M} \sum e^{-i
\theta_i} \frac{v(i)}{|v(i)|} \right|$.  For the template methods,
shown are the quantities $\rho = \left|\frac{1}{M} \sum e^{-i
\theta_i} \mbox{exp}(2 i \pi a_i / 128) \right|$, where the
$\theta_i$ are the true alignments and the $a_i$ are the alignments
predicted by various templates:  a cosine, the second moment, a
triangle, the Mexican hat, the second Haar wavelet, a noiseless MTW
snapshot (use of it is ``cheating''), and a noisy MTW snapshot (not
``cheating'').  For a fixed amount of noise, $\lambda_H$ appears to
increase approximately with $M$, while $\lambda_R$ increases
approximately with $\sqrt{M}$.  As the amount of noise increases
($\sigma^2$), $\lambda_H$ decreases, $\lambda_R$ increases, and, as
expected, both of the $\rho_\lambda$ decrease (this is expected both
intuitively and mathematically, see Section
\ref{sec:SummaryMethod}). Similarly, other template-based
correlations $\rho$ increase with $M$ and decrease with decreasing
noise. Clearly, the only competitive template is a noiseless
snapshot of the MTW itself.\label{table:KSE}}
\end{center}
\end{table}

\subsection{Additional denoising procedures}

Before concluding this example, we note that if we initially filter
the noisy wave snapshots, we observe better performance for both the
eigenvector method and for some of the fixed templates.
In the Fourier representation of the {\em non-noisy} KSE snapshots
(convenient for spectral numerical discretization, but also known to
be the optimal principal component (PCA)  basis for systems with
such translational symmetry, see \cite{fourOpt}) the power spectrum
is known to decay quickly.
Therefore, we obtain an increased signal-to-noise ratio by first
projecting each noisy wave snapshot onto its Fourier modes with
power spectrua larger than some fixed threshold; the
\textit{information} about the underlying (non-noisy) MTW attractor
which is thrown away by filtering these Fourier modes is small
compared to the \textit{noise} thrown away by filtering these
Fourier modes.
%


\section{Post-processing the aligned data of the Kuramoto-Sivashinsky equation through the use of
diffusion maps} \label{sec:DMAPS}

In the example of the KSE wave snapshots (Section \ref{sec:KSE}), we
conveniently allowed ourselves to ignore the shape modulation
superposed to the traveling motion when seeking their global
alignments.
The reason is that this modulation is comparatively small in $L^2$
norm, and therefore, it contributes little to the sum in
equation (\ref{eq:maxCorr}).
We were able to recover, with
good accuracy, the global alignments of the noisy wave snapshots (see
Figure \ref{fig:synch}).

\begin{figure}
\begin{center} \includegraphics[width=125mm]{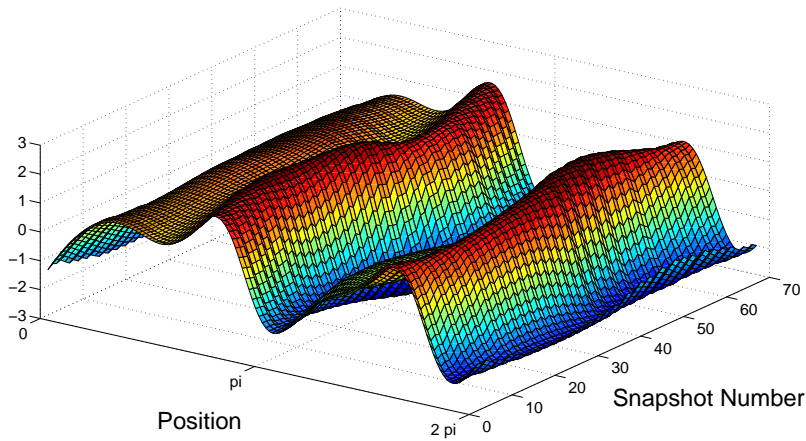} \end{center}
    \caption{A sequence of \textit{aligned} MTW snapshots.  In contrast to Figure \ref{fig:MTWseq}, the traveling has been factored out and only the modulation remains.  Although the eigenvector method performs well on noisy snapshots (see, e.g., Figure \ref{fig:synch}), we chose to show noise-free MTW snapshots for visualization purposes.}
    \label{fig:MTWseqAlign}
\end{figure}

With the global alignments recovered, we rotate each snapshot so
that the traveling motion is factored out and only the modulation
remains.
When there is no noise, the aligned sequence of wave
snapshots takes the form of Figure \ref{fig:MTWseqAlign} (with
noise it is too hard to visually perceive the modulation, so we do not include such
a figure).

Given the aligned data, we now perform diffusion maps in order to
search for ``coarse variables" (that is, for reduced representations
of the data) as in
\cite{lafonlee,physreve,cecilia,coifman2,frewenchem}).
To construct an informative
low-dimensional embedding for this data set of $M$ (noisy but
aligned) snapshots, we start with a similarity measure between each
pair of snapshots $u(x,t_i)$, $u(x,t_j)$.
The similarity measure is a nonnegative quantity $\mathbf{W}_{ij} =
\mathbf{W}_{ji}$ satisfying certain additional ``admissibility
conditions" (\cite{coifman1}).
Here, we choose the Gaussian similarity measure, and construct a matrix $\mathbf{W}$ as
\begin{equation}
 \mathbf{W}_{ij} = \mbox{exp}\left[ \frac{\sum_{k=1}^{128} \left[ u(x_k,t_i)-u(x_k,t_j) \right]^2 }{\varepsilon} \right].
\end{equation}
In this equation, $\varepsilon$ defines a characteristic scale which
quantifies the ``locality'' of the neighborhood within which
Euclidean distance can be used as the basis of a meaningful
similarity measure (\cite{coifman1}).
A systematic approach to determining appropriate $\varepsilon$
values is discussed in \cite{loglog}.
Next, we create a matrix $\mathbf{K}$ which is a row-normalized version of $\mathbf{W}$:
\begin{equation}
 \mathbf{K}_{ij} = \frac{\mathbf{W}_{ij}}{\sum_{l=1}^M \mathbf{W}_{il}}.
\end{equation}
Finally, we look at the top few eigenvalues and eigenvectors of the matrix $\mathbf{K}$.
In MATLAB, for instance, this can be done with the command
$[\mathbf{V},\mathbf{L}] = \mbox{eigs}(\mathbf{K},n+1)$, where $n+1$
is the number of top eigenvalues we wish to keep (we typically are
only interested in the first few).

This gives a set of real eigenvalues $\lambda_0 \geq \lambda_1 \geq
... \geq \lambda_{n} \geq 0$ with corresponding eigenvectors $\{\vec
\psi_j\}_{j=0}^{n}$.
Since $\mathbf{K}$ is stochastic, $\lambda_0 = 1$ and $\vec \psi_0 =
[1\, 1 \, ... \, 1]^T$.
The $n$-dimensional representation of the $i$-th snapshot $u(x,t_i)$ is given by the \textit{diffusion map}
$\vec \Psi_n^{(i)}:\, \mathbf{R}^{128} \longrightarrow \mathbf{R}^n$, where
\begin{equation*}
\vec \Psi_n\left(u(x,t_i)\right) = [\lambda_1^t \vec \psi_1^{(i)}, \lambda_2^t \vec \psi_2^{(i)}, ..., \lambda_n^t \vec
\psi_n^{(i)}],
\end{equation*}
a mapping which is only defined on the $M$ recorded snapshots.
Here, $t$ represents the ``diffusion time''; to keep things simple, we choose $t=1$.
In other words, snapshot $i$ is mapped to a vector whose
first component is the $i$th component of the first nontrivial
eigenvector, whose second component is the $i$th component of the
second nontrivial eigenvector, etc.
If a gap in the eigenvalue spectrum is observed between eigenvalues
$\lambda_n$ and $\lambda_{n+1}$, then $\vec \Psi_n$ may provide a
useful low-dimensional representation of the data set
(\cite{belkin2003,coiffp}).

\begin{figure}
\begin{center} \includegraphics[width=125mm]{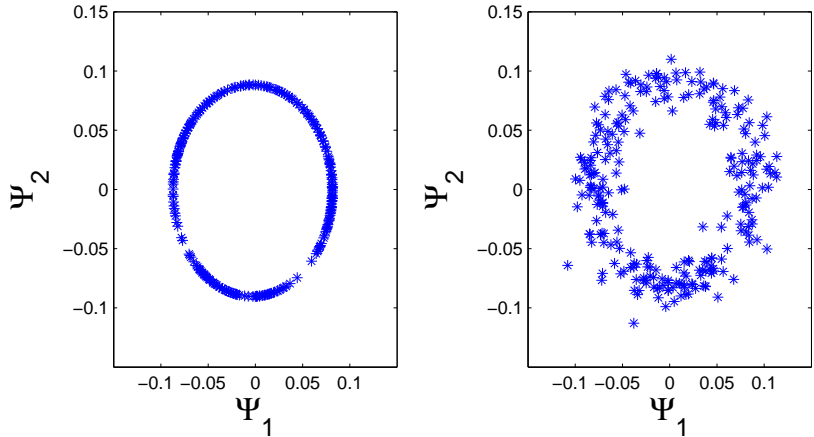} \end{center}
    \caption{The two diffusion map coordinates (the first two nontrivial eigenvectors of $\mathbf{K}$) obtained from aligned, but noisy, MTW snapshots (right) and aligned, noise-free MTW snapshots (left).}
    \label{fig:eVecsKSE}
\end{figure}

\begin{figure}
\begin{center} \includegraphics[width=125mm]{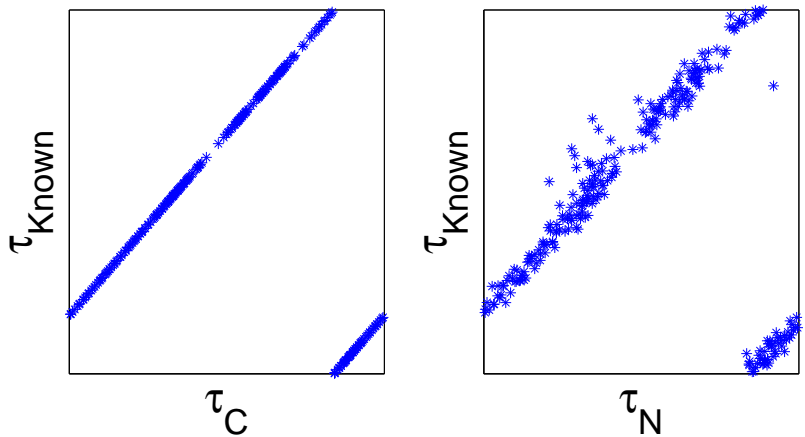} \end{center}
    \caption{Parametrizations of the modulations via diffusion maps.  On the $x$ axes, $\tau_C$ and $\tau_N$ (``clean'' and ''noisy,'' respectively), are computed from the diffusion map
eigenvector information in Figure \ref{fig:eVecsKSE}.  On the $y$
axes, $\tau_{Known}$ is the ``correct'' parametrization. The plot of
$\tau_{Known}$ vs. $\tau_C$ is shown just for comparison, for
$\tau_C$ and $\tau_{Known}$ differ by only a phase offset. The two
figures are roughly one-to-one (modulo $2 \pi$).}
    \label{fig:modComp}
\end{figure}

When we apply diffusion maps to the (aligned but noisy) wave snapshot
data, our eigenvalues are $1.00,\, 0.90,\, 0.87,\, 0.62,\, 0.43,\,
\hdots$
Clearly, there is a gap between $0.87$ and $0.62$.  Therefore, we
expect the first two nontrivial eigenvectors to give a
parametrization of the residual, ``symmetry-adjusted" dynamics
corresponding to the modulation.
These two eigenvectors are shown in Figure
\ref{fig:eVecsKSE}.
There \textit{is} a continuous one-to-one map
between each possible modulation phase and the set of points on the
unit circle, since the data lie very close to the attracting
modulated traveling wave, for which the modulation is exactly
periodic in time.
%
We thus expect the first two nontrivial eigenvectors to trace out
some sort of circle or ``loop''; the eigenfunctions of simple
diffusion on a closed curve are $\sin (2 s \pi / L)$ and $\cos (2 s
\pi / L)$, where $s$ is some arclength parameter.
The eigenvectors shown in
Figure \ref{fig:eVecsKSE} do not trace out an exact circle,
but the plot is reminiscent of that  shape.
In fact, by looking at
the quantity
\begin{equation}
 \tau_i \equiv \mbox{arctan} \left( \frac{\vec \psi_2^{(i)}}{\vec \psi_1^{(i)}} \right),
\end{equation}
we can assign a number $\tau_i \in [0,2 \pi)$ to each snapshot,
parameterizing the modulation.
When we plot $\tau$ against a
known parametrization of the modulation, we obtain Figure
\ref{fig:modComp}.
As the two quantities are approximately one-to-one (modulo $2 \pi$),
it is clear that our diffusion map analysis has been successful in
parameterizing the modulation, the residual dynamics of the
symmetry-adjusted snapshots.
Given the small size of the modulations compared to the
overall noise of the problem, this is encouraging.

\section{Vector diffusion maps} \label{sec:vDMAPS}

In the preceding sections, we were able to take advantage of the
eigenvector alignment method to provide information about the global
alignment of ensembles of snapshots in two illustrative pattern-forming
systems with symmetry.
In the case of the orientational probability distributions of
nematic liquid crystals (Section \ref{sec:LCP}), all $M$ snapshots
were in principle rotated versions of the \textit{same} distribution
function; due to the finiteness of the representation, however (each
was a collection of $N$ representative particles), noise became a
feature of the problem.
For the spatiotemporally varying wave snapshots of the
Kuramoto-Sivashinksy equation (Section \ref{sec:KSE}), we were able
to apply the eigenvector method to factor out the traveling
component of the variation, even though each snapshot was
\textit{not} exactly the same up to rotation.
We were successful because the modulation (superposed to the traveling component of
the evolution) was relatively small.
We then applied diffusion maps to the aligned
snapshots and successfully recovered a meaningful, low-dimensional
representation of the residual dynamics (the modulation).

Now, suppose that in the case of the LCP orientational probability
distributions, the set of $M$ snapshots contained not only rotated
realizations of the same (noisy) distribution, but also randomly
rotated versions of snapshots that had evolved for different lengths
of time.
The differences in the $M$ snapshots would then be due to
\begin{itemize}
\item different finite particle realizations of the distribution function (only in the case $N \rightarrow \infty$ do they become the same);
\item different rotations of these distribution function realizations; and
\item the fact that the distribution function changes with time.
\end{itemize}
In such a situation, it might not be prudent to try to align two
orientational probability distribution functions of vastly different
shapes (these may arise from evolution over appreciably different
lengths of time).
A similar situation might arise if the modulation in our traveling/modulating
wave snapshots is \textit{not}
small: pairwise alignments of vastly different shapes would stop being meaningful.
{\em Vector Diffusion Maps} (\cite{vdm}) provide an approach that,
in such circumstances, both help obtain global alignments and also
reveal the underlying ``symmetry-adjusted'' reduced dynamics all in
one step.

\subsection{A brief introduction to vector diffusion maps}

The reduced descriptions of the dynamics obtained by diffusion maps
(as we did in the KSE example above)
rely on the user's ability to provide a pairwise similarity measure
$\mathbf{W}_{ij}$ between snapshots $i$ and $j$.
From there, the
largest eigenvalues (and corresponding eigenvectors) of a matrix
$\mathbf{K}$ are computed, where $\mathbf{K}_{ij} = \mathbf{W}_{ij}
/ \sum_k \mathbf{W}_{ik}$.
In the case of the KSE wave snapshots, we set
\begin{equation}
 \mathbf{W}_{ij} = \mbox{exp}\left[ \frac{\sum_{k=1}^{128} \left[ u(x_k,t_i)-u(x_k,t_j) \right]^2 }{\varepsilon} \right]
\end{equation}
(see Section \ref{sec:DMAPS}).
Intuitively, the eigenvectors of $\mathbf{K}$ corresponding to the
largest eigenvalues are those related to the most robust diffusions
in a graph whose vertices are the data (see, e.g.
\cite{belkin2003}); if snapshot $i$ is ``close'' to snapshot $j$ in
diffusion map space, then it should be possible to transition from
the one to the other easily through mutually neighboring snapshots
$k$, neighbors of neighbors, etc.

Likewise, the global alignments provided by the eigenvector method
rely on the user to first compare all snapshots in a pairwise
fashion so as to obtain the group element $g_{ij} \in G$ which
``best'' aligns them, and then incorporate the real/complex
representation of this group element, $\mathbf{O}_{ij}$, into the
$ij$-th block of a matrix.
In the case of the KSE wave
snapshots, we denoted this group element $\mathbf{O}_{ij}$ as
\begin{equation}
\mathbf{T}_{ij} = \mbox{exp}(-2i \pi a_{ij}/128)
\end{equation}
(see Section \ref{sec:SummaryMethod}).
Intuitively, the eigenvector of $\mathbf{O}$ with largest
corresponding eigenvalue corresponds to the most consistent global
alignment; if snapshot $i$ can be rotated to snapshot $j$ via
$g_{ij}$, then snapshot $i$ should also be able to be rotated to
snapshot $j$ through a snapshot $k$ (via $g_{ik} g_{kj}$).

Vector diffusion maps attempts, in a sense, to combine the two
methods (the eigenvector method and diffusion maps).
To use vector diffusion maps, one first optimally aligns
two snapshots $i$ and $j$ to obtain $g_{ij}$ and thus
$\mathbf{O}_{ij}$; one then computes the similarity of $i$ and $j$ after
this alignment has taken place to obtain $\mathbf{W}_{ij}$ (and,
after normalization, $\mathbf{K}_{ij}$).
A matrix $\mathbf{S}$ is
then formed whose $ij$-th block is simply $\mathbf{S}_{ij} =
\mathbf{K}_{ij} \mathbf{O}_{ij}$.
The eigenvectors of $\mathbf{S}$ corresponding to its largest
eigenvalues are computed, and these eigenvectors provide information
about both symmetry adjustment (``alignment") {\em and} about
dynamic similarity.
Distances between snapshots in this new vector diffusion map space
are called \textit{vector diffusion distances} (see equations (4.2)
and (4.6) on p. 11 of \cite{vdm}).
As we noted above, alignment comparisons between snapshots should
only be trusted when $\mathbf{W}_{ij}$ is not small, for it may not
make sense to compare two snapshots which differ appreciably (e.g.
in shape and/or in temporal evolution time ).
Vector diffusion maps accomplishes this
by effectively ignoring comparisons $\mathbf{O}_{ij}$ for snapshots
which are ``far away'' (small $\mathbf{W}_{ij}$) from each other.

\subsection{Application of vector diffusion maps to the
spatiotemporal wave snapshots of the KSE}


To apply vector diffusion maps to the KSE example, we form the matrix
$\mathbf{S}$ by setting
\begin{equation}
\mathbf{S}_{ij} =  \mathbf{T}_{ij} \mathbf{K}_{ij},
\end{equation}
where the the $\mathbf{T}_{ij}$ are obtained by optimally aligning
each pair of noisy wave snapshots, and the $\mathbf{K}_{ij}$ are
then computed on the symmetry-adjusted wave snapshots (these
$\mathbf{K}_{ij}$ are, as before, $\mathbf{W}_{ij} / \sum_k
\mathbf{W}_{ik} $).

The top eigenvectors of $\mathbf{S}$ are then computed, and the
eigenvalues are exactly as in Section \ref{sec:DMAPS}:  $1.00,\,
0.90,\, 0.87,\, 0.62,\, 0.43,\, \hdots$
This is not surprising, for this particular problem actually
``decouples''; the modulation is independent of the traveling motion
for an exact modulated traveling wave (in other, more general
problems, this is unlikely to be the case).
The first eigenvector $\mathbf{v}_0$, the one corresponding to
eigenvalue $1.00$, reveals the global alignments (see Section
\ref{sec:KSEeVec}) and has the form
\begin{equation}
\mathbf{v}_0 = \left[\begin{array}{c} \mbox{exp}(2i\pi a_1/128) \\
\mbox{exp}(2i\pi a_2/128) \\ \vdots \\ \mbox{exp}(2i\pi a_M/128)
\end{array} \right].
\end{equation}
The next two eigenvectors reveal the diffusion map parametrization
of the underlying, symmetry-adjusted dynamics (the modulation, see
Section \ref{sec:DMAPS}).
These eigenvectors are ``corrupted'' because they also contain the
global alignments:
\begin{equation}
\mathbf{v}_{1,2} = \left[\begin{array}{c} \mbox{exp}(2i\pi a_1/128) \vec \psi_{1,2}^{(1)} \\
\mbox{exp}(2i\pi a_2/128) \vec \psi_{1,2}^{(2)} \\ \vdots \\
\mbox{exp}(2i\pi a_M/128) \vec \psi_{1,2}^{(M)}
\end{array} \right].
\end{equation}
However, one can easily get $\mathbf{v}_1$ and $\mathbf{v}_2$ back
to their more meaningful, ``original'' forms $\vec \psi_1$ and $\vec
\psi_2$ of Section \ref{sec:DMAPS} by simply dividing each of them
entrywise by $\mathbf{v}$.

Since this particular example ``decouples,'' the global alignments and
eigenvectors obtained this way agree with those already computed and shown in Figures
\ref{fig:synch} and \ref{fig:eVecsKSE}; we do not plot them again here.

\section{Summary and conclusions}

In this paper we applied both the ``eigenvector method"
(\cite{amit,cryo1,cryo2}), and vector diffusion maps (\cite{vdm})
(based on the eigenvector method) to adjust data ensembles
(consisting of snapshots from two evolving systems) with respect to
the system intrinsic symmetries.
We demonstrated the ability of both vector diffusion maps and the
eigenvector method to align (and in a sense, denoise) the data sets,
and also parameterize their symmetry-adjusted dynamics.
For both examples, the
eigenvector method provided a global alignment of the noisy
snapshots of the evolving systems, even with a small
signal-to-noise ratio.
Additionally, for the case of
traveling and modulating waves, vector diffusion maps were shown to both remove the
underlying symmetry and capture the underlying long-term dynamics (the
residual dynamics of modulation, after the ``traveling'' symmetry has
been removed).

The two techniques are fast and easy to implement, and as discussed,
they are a natural analogue to diffusion maps
(\cite{coifman_diffusion}) in the sense that they rely on pairwise
comparison data.
This information is incorporated into an eigenvalue problem, whose
result is a globally consistent (in a certain sense, see Section
\ref{sec:SummaryMethod}) parametrization/alignment of the underlying
data set.
Just as diffusion maps are robust to noise in the
computation of the pairwise similarity measurements, vector
diffusion maps and the eigenvector method are robust to both noise
and alignment error in the computation of both the pairwise
similarity measurements and symmetry group members.

By taking into account the equivariance of the system dynamics with
respect to the underlying symmetry, vector diffusion maps may reduce the amount
of data required in order to successfully elucidate an effective,
low-dimensional description of the dynamics.
Despite the success of
nonlinear dimensionality reduction techniques in finding meaningful
reduced descriptions for complex systems (see, e.g.
\cite{frewenchem,physreve,cecilia}), they still suffer from the
curse of dimensionality; in general, the amount of data required to
successfully recover $d$ ``intrinsic'' dimensions grows
exponentially with $d$.
Factoring out dimensions associated with the symmetry degrees of
freedom will partially alleviate of this problem.  While diffusion
maps treats the snapshots as living on a manifold $\mathcal{M}$,
vector diffusion maps in effect treats the snapshots as if they live
in the quotient space $\mathcal{M} / G$. This implicit reduction of
dimensionality allows the methods presented in this paper to provide
an improved organization of the data.

\section{Acknowledgments}

B.E.S. was partially supported by the DOE CSGF (grant number
DE-FG02-97ER25308) and the NSF GRFP.  A.S. and I.G.K. were partially
supported by the DOE (grant numbers DE-SC0002097 and DE-SC0005176),
and A.S. also thanks the Sloan research fellowship. The authors
would also like to acknowledge Constantinos I. Siettos for
generously providing the LCP codes used in Section \ref{sec:LCP}.

\bibliographystyle{model5-names}
\bibliography{Templstuff_new}

\begin{thebibliography}{48}
\expandafter\ifx\csname natexlab\endcsname\relax\def\natexlab#1{#1}\fi
\providecommand{\bibinfo}[2]{#2}
\ifx\xfnm\relax \def\xfnm[#1]{\unskip,\space#1}\fi
\bibitem[{Ahuja et~al.(2007)Ahuja, Kevrekidis \& Rowley}]{rowley1}
\bibinfo{author}{Ahuja, S.}, \bibinfo{author}{Kevrekidis, I.~G.}, \&
  \bibinfo{author}{Rowley, C.~W.} (\bibinfo{year}{2007}).
\newblock \bibinfo{title}{{Template-based stabilization of relative equilibria
  in systems with continuous symmetry}}.
\newblock {\it \bibinfo{journal}{Journal of Nonlinear Science}\/},  {\it
  \bibinfo{volume}{17}\/}, \bibinfo{pages}{109--143}.
\bibitem[{Arecchi et~al.(1999)Arecchi, Boccaletti \& Ramazza}]{partexamp2}
\bibinfo{author}{Arecchi, F.~T.}, \bibinfo{author}{Boccaletti, S.}, \&
  \bibinfo{author}{Ramazza, P.~L.} (\bibinfo{year}{1999}).
\newblock \bibinfo{title}{{Pattern formation and competition in nonlinear
  optics}}.
\newblock {\it \bibinfo{journal}{Physics Reports}\/},  {\it
  \bibinfo{volume}{318}\/}, \bibinfo{pages}{1--83}.
\bibitem[{Aubry et~al.(1993)Aubry, Lian \& Titi}]{pres_symm_pod}
\bibinfo{author}{Aubry, N.}, \bibinfo{author}{Lian, W.}, \&
  \bibinfo{author}{Titi, E.} (\bibinfo{year}{1993}).
\newblock \bibinfo{title}{{Preserving symmetries in the proper orthogonal
  decomposition}}.
\newblock {\it \bibinfo{journal}{SIAM Journal on Scientific Computing}\/},
  {\it \bibinfo{volume}{14}\/}, \bibinfo{pages}{483}.
\bibitem[{Belkin \& Niyogi(2003)}]{belkin2003}
\bibinfo{author}{Belkin, M.}, \& \bibinfo{author}{Niyogi, P.}
  (\bibinfo{year}{2003}).
\newblock \bibinfo{title}{{Laplacian eigenmaps for dimensionality reduction and
  data representation}}.
\newblock {\it \bibinfo{journal}{Neural computation}\/},  {\it
  \bibinfo{volume}{15}\/}, \bibinfo{pages}{1373--1396}.
\bibitem[{Berkooz et~al.(1993)Berkooz, Holmes \& Lumley}]{podg2}
\bibinfo{author}{Berkooz, G.}, \bibinfo{author}{Holmes, P.}, \&
  \bibinfo{author}{Lumley, J.~L.} (\bibinfo{year}{1993}).
\newblock \bibinfo{title}{{The proper orthogonal decomposition in the analysis
  of turbulent flows}}.
\newblock {\it \bibinfo{journal}{Annual Review of Fluid Mechanics}\/},  {\it
  \bibinfo{volume}{25}\/}, \bibinfo{pages}{539--575}.
\bibitem[{Berkooz \& Titi(1993)}]{podg3}
\bibinfo{author}{Berkooz, G.}, \& \bibinfo{author}{Titi, E.~S.}
  (\bibinfo{year}{1993}).
\newblock \bibinfo{title}{{Galerkin projections and the proper orthogonal
  decomposition for equivariant equations}}.
\newblock {\it \bibinfo{journal}{Physics Letters A}\/},  {\it
  \bibinfo{volume}{174}\/}, \bibinfo{pages}{94--102}.
\bibitem[{Cheng \& Singer()}]{Xiuyuan}
\bibinfo{author}{Cheng, X.}, \& \bibinfo{author}{Singer, A.} ().
\newblock \bibinfo{title}{{The Spectrum of an Hermitian Matrix With Dependent
  Entries Constructed from Random Independent Images}}.
\newblock {\it \bibinfo{journal}{in preparation}\/}, .
\bibitem[{Coifman et~al.(2005{\natexlab{a}})Coifman, Lafon, Lee, Maggioni,
  Nadler, Warner \& Zucker}]{coifman1}
\bibinfo{author}{Coifman, R.}, \bibinfo{author}{Lafon, S.},
  \bibinfo{author}{Lee, A.}, \bibinfo{author}{Maggioni, M.},
  \bibinfo{author}{Nadler, B.}, \bibinfo{author}{Warner, F.}, \&
  \bibinfo{author}{Zucker, S.} (\bibinfo{year}{2005}{\natexlab{a}}).
\newblock \bibinfo{title}{Geometric diffusions as a tool for harmonic analysis
  and structure definition of data: Diffusion maps}.
\newblock {\it \bibinfo{journal}{PNAS}\/},  {\it \bibinfo{volume}{102}\/},
  \bibinfo{pages}{7426}.
\bibitem[{Coifman et~al.(2005{\natexlab{b}})Coifman, Lafon, Lee, Maggioni,
  Nadler, Warner \& Zucker}]{coifman2}
\bibinfo{author}{Coifman, R.}, \bibinfo{author}{Lafon, S.},
  \bibinfo{author}{Lee, A.}, \bibinfo{author}{Maggioni, M.},
  \bibinfo{author}{Nadler, B.}, \bibinfo{author}{Warner, F.}, \&
  \bibinfo{author}{Zucker, S.} (\bibinfo{year}{2005}{\natexlab{b}}).
\newblock \bibinfo{title}{Geometric diffusions as a tool for harmonic analysis
  and structure definition of data: Multiscale methods}.
\newblock {\it \bibinfo{journal}{PNAS}\/},  {\it \bibinfo{volume}{102}\/},
  \bibinfo{pages}{7432}.
\bibitem[{Coifman \& Lafon(2006)}]{coifman_diffusion}
\bibinfo{author}{Coifman, R.~R.}, \& \bibinfo{author}{Lafon, S.}
  (\bibinfo{year}{2006}).
\newblock \bibinfo{title}{{Diffusion maps}}.
\newblock {\it \bibinfo{journal}{Applied and Computational Harmonic
  Analysis}\/},  {\it \bibinfo{volume}{21}\/}, \bibinfo{pages}{5--30}.
\bibitem[{Constantin et~al.(1988)Constantin, Foias, Nicolaenko \&
  Temam}]{initmanbook}
\bibinfo{author}{Constantin, P.}, \bibinfo{author}{Foias, C.},
  \bibinfo{author}{Nicolaenko, B.}, \& \bibinfo{author}{Temam, R.}
  (\bibinfo{year}{1988}).
\newblock {\it \bibinfo{title}{Integral manifolds and inertial manifolds for
  dissipative partial differential equations}\/}.
\newblock \bibinfo{publisher}{``Applied Mathematical Science Series,'' No. 70,
  Springer-Verlag, New York}.
\bibitem[{Cross \& Hohenberg(1993)}]{partexamp1}
\bibinfo{author}{Cross, M.~C.}, \& \bibinfo{author}{Hohenberg, P.~C.}
  (\bibinfo{year}{1993}).
\newblock \bibinfo{title}{{Pattern formation outside of equilibrium}}.
\newblock {\it \bibinfo{journal}{Reviews of Modern Physics}\/},  {\it
  \bibinfo{volume}{65}\/}, \bibinfo{pages}{851--1112}.
\bibitem[{Das et~al.(2006)Das, Moll, Stamati, Kavraki \& Clementi}]{cecilia}
\bibinfo{author}{Das, P.}, \bibinfo{author}{Moll, M.},
  \bibinfo{author}{Stamati, H.}, \bibinfo{author}{Kavraki, L.}, \&
  \bibinfo{author}{Clementi, C.} (\bibinfo{year}{2006}).
\newblock \bibinfo{title}{Low-dimensional, free-energy landscapes of
  protein-folding reactions by nonlinear dimensionality reduction}.
\newblock {\it \bibinfo{journal}{PNAS}\/},  {\it \bibinfo{volume}{103}\/},
  \bibinfo{pages}{9885}.
\bibitem[{Erban et~al.(2007)Erban, Frewen, Wang, Elston, Coifman, Nadler \&
  Kevrekidis}]{frewenchem}
\bibinfo{author}{Erban, R.}, \bibinfo{author}{Frewen, T.~A.},
  \bibinfo{author}{Wang, X.}, \bibinfo{author}{Elston, T.~C.},
  \bibinfo{author}{Coifman, R.}, \bibinfo{author}{Nadler, B.}, \&
  \bibinfo{author}{Kevrekidis, I.~G.} (\bibinfo{year}{2007}).
\newblock \bibinfo{title}{{Variable-free exploration of stochastic models: a
  gene regulatory network example}}.
\newblock {\it \bibinfo{journal}{The Journal of chemical physics}\/},  {\it
  \bibinfo{volume}{126}\/}, \bibinfo{pages}{155103}.
\bibitem[{Fan \& Hoffman(1955)}]{Fan1955}
\bibinfo{author}{Fan, K.}, \& \bibinfo{author}{Hoffman, A.~J.}
  (\bibinfo{year}{1955}).
\newblock \bibinfo{title}{Some metric inequalities in the space of matrices}.
\newblock {\it \bibinfo{journal}{Proceedings of the American Mathematical
  Society}\/},  {\it \bibinfo{volume}{6}\/}, \bibinfo{pages}{111--116}.
\bibitem[{F{\'e}ral \& P{\'e}ch{\'e}(2007)}]{feral}
\bibinfo{author}{F{\'e}ral, D.}, \& \bibinfo{author}{P{\'e}ch{\'e}, S.}
  (\bibinfo{year}{2007}).
\newblock \bibinfo{title}{{The largest eigenvalue of rank one deformation of
  large Wigner matrices}}.
\newblock {\it \bibinfo{journal}{Communications in Mathematical Physics}\/},
  {\it \bibinfo{volume}{272}\/}, \bibinfo{pages}{185--228}.
\bibitem[{Foias et~al.(1988{\natexlab{a}})Foias, Jolly, Kevrekidis, Sell \&
  Titi}]{aimshort}
\bibinfo{author}{Foias, C.}, \bibinfo{author}{Jolly, M.~S.},
  \bibinfo{author}{Kevrekidis, I.~G.}, \bibinfo{author}{Sell, G.~R.}, \&
  \bibinfo{author}{Titi, E.~S.} (\bibinfo{year}{1988}{\natexlab{a}}).
\newblock \bibinfo{title}{{On the computation of inertial manifolds}}.
\newblock {\it \bibinfo{journal}{Physics Letters A}\/},  {\it
  \bibinfo{volume}{131}\/}, \bibinfo{pages}{433--436}.
\bibitem[{Foias et~al.(1988{\natexlab{b}})Foias, Sell \&
  Temam}]{foias1988inertial}
\bibinfo{author}{Foias, C.}, \bibinfo{author}{Sell, G.~R.}, \&
  \bibinfo{author}{Temam, R.} (\bibinfo{year}{1988}{\natexlab{b}}).
\newblock \bibinfo{title}{{Inertial manifolds for nonlinear evolutionary
  equations}}.
\newblock {\it \bibinfo{journal}{Journal of Differential Equations}\/},  {\it
  \bibinfo{volume}{73}\/}, \bibinfo{pages}{309--353}.
\bibitem[{Foias et~al.(1989)Foias, Sell \& Titi}]{foias1989exponential}
\bibinfo{author}{Foias, C.}, \bibinfo{author}{Sell, G.~R.}, \&
  \bibinfo{author}{Titi, E.~S.} (\bibinfo{year}{1989}).
\newblock \bibinfo{title}{{Exponential tracking and approximation of inertial
  manifolds for dissipative nonlinear equations}}.
\newblock {\it \bibinfo{journal}{Journal of Dynamics and Differential
  Equations}\/},  {\it \bibinfo{volume}{1}\/}, \bibinfo{pages}{199--244}.
\bibitem[{Grassberger \& Procaccia(1983)}]{loglog}
\bibinfo{author}{Grassberger, P.}, \& \bibinfo{author}{Procaccia, I.}
  (\bibinfo{year}{1983}).
\newblock \bibinfo{title}{{Measuring the strangeness of strange attractors}}.
\newblock {\it \bibinfo{journal}{Physica D: Nonlinear Phenomena}\/},  {\it
  \bibinfo{volume}{9}\/}, \bibinfo{pages}{189--208}.
\bibitem[{Guckenheimer \& Holmes(2002)}]{guckholmesbook}
\bibinfo{author}{Guckenheimer, J.}, \& \bibinfo{author}{Holmes, P.}
  (\bibinfo{year}{2002}).
\newblock {\it \bibinfo{title}{{Nonlinear oscillations, dynamical systems, and
  bifurcations of vector fields}}\/}.
\newblock \bibinfo{publisher}{Springer}.
\bibitem[{Holmes et~al.(1998)Holmes, Lumley \& Berkooz}]{holmes1998turbulence}
\bibinfo{author}{Holmes, P.}, \bibinfo{author}{Lumley, J.~L.}, \&
  \bibinfo{author}{Berkooz, G.} (\bibinfo{year}{1998}).
\newblock {\it \bibinfo{title}{{Turbulence, coherent structures, dynamical
  systems and symmetry}}\/}.
\newblock \bibinfo{publisher}{Cambridge Univ Pr}.
\bibitem[{Jolly(1989)}]{jolly1989explicit}
\bibinfo{author}{Jolly, M.~S.} (\bibinfo{year}{1989}).
\newblock \bibinfo{title}{{Explicit construction of an inertial manifold for a
  reaction diffusion equation}}.
\newblock {\it \bibinfo{journal}{Journal of Differential Equations}\/},  {\it
  \bibinfo{volume}{78}\/}, \bibinfo{pages}{220--261}.
\bibitem[{Jolly et~al.(1990)Jolly, Kevrekidis \& Titi}]{aim}
\bibinfo{author}{Jolly, M.~S.}, \bibinfo{author}{Kevrekidis, I.~G.}, \&
  \bibinfo{author}{Titi, E.~S.} (\bibinfo{year}{1990}).
\newblock \bibinfo{title}{{Approximate inertial manifolds for the
  Kuramoto-Sivashinsky equation: analysis and computations}}.
\newblock {\it \bibinfo{journal}{Physica D}\/},  {\it \bibinfo{volume}{44}\/},
  \bibinfo{pages}{38--60}.
\bibitem[{Keller(1975)}]{Keller1975}
\bibinfo{author}{Keller, J.~B.} (\bibinfo{year}{1975}).
\newblock \bibinfo{title}{Closest unitary, orthogonal and hermitian operators
  to a given operator}.
\newblock {\it \bibinfo{journal}{Mathematics Magazine}\/},  {\it
  \bibinfo{volume}{48}\/}, \bibinfo{pages}{192--197}.
\bibitem[{Kevrekidis et~al.(2008)Kevrekidis, Frantzeskakis \&
  Carretero-Gonz{\'a}lez}]{partexamp3}
\bibinfo{author}{Kevrekidis, P.~G.}, \bibinfo{author}{Frantzeskakis, D.~J.}, \&
  \bibinfo{author}{Carretero-Gonz{\'a}lez, R.} (\bibinfo{year}{2008}).
\newblock \bibinfo{title}{{Emergent nonlinear phenomena in Bose-Einstein
  condensates}}.
\newblock {\it \bibinfo{journal}{Theory and experiment}\/}, .
\bibitem[{Kunisch \& Volkwein(2003)}]{podg1}
\bibinfo{author}{Kunisch, K.}, \& \bibinfo{author}{Volkwein, S.}
  (\bibinfo{year}{2003}).
\newblock \bibinfo{title}{{Galerkin proper orthogonal decomposition methods for
  a general equation in fluid dynamics}}.
\newblock {\it \bibinfo{journal}{SIAM Journal on Numerical Analysis}\/},  (pp.
  \bibinfo{pages}{492--515}).
\bibitem[{Kuramoto \& Tsuzuki(1976)}]{ksfilm}
\bibinfo{author}{Kuramoto, Y.}, \& \bibinfo{author}{Tsuzuki, T.}
  (\bibinfo{year}{1976}).
\newblock \bibinfo{title}{{Persistent propagation of concentration waves in
  dissipative media far from thermal equilibrium}}.
\newblock {\it \bibinfo{journal}{Prog. Theor. Phys}\/},  {\it
  \bibinfo{volume}{55}\/}, \bibinfo{pages}{356--369}.
\bibitem[{Lafon \& Lee(2006)}]{lafonlee}
\bibinfo{author}{Lafon, S.}, \& \bibinfo{author}{Lee, A.~B.}
  (\bibinfo{year}{2006}).
\newblock \bibinfo{title}{Diffusion maps and coarse-graining: a unified
  framework for dimensionality reduction, graph partitioning, and data set
  parameterization}.
\newblock {\it \bibinfo{journal}{Pattern Analysis and Machine Intelligence,
  IEEE Transactions on}\/},  {\it \bibinfo{volume}{28}\/},
  \bibinfo{pages}{1393}.
\bibitem[{Maier \& Saupe(1959)}]{maiersaupe}
\bibinfo{author}{Maier, W.}, \& \bibinfo{author}{Saupe, A.}
  (\bibinfo{year}{1959}).
\newblock \bibinfo{title}{{Eine einfache molekularstatistische Theorie der
  nematischen kristallinfl{\\"u}ssigen Phase. Teil I}}.
\newblock {\it \bibinfo{journal}{Zeitschrift Naturforschung Teil A}\/},  {\it
  \bibinfo{volume}{14}\/}, \bibinfo{pages}{882}.
\bibitem[{Metropolis et~al.(1953)Metropolis, Rosenbluth, Rosenbluth, Teller \&
  Teller}]{metropolis}
\bibinfo{author}{Metropolis, N.}, \bibinfo{author}{Rosenbluth, A.~W.},
  \bibinfo{author}{Rosenbluth, M.~N.}, \bibinfo{author}{Teller, A.~H.}, \&
  \bibinfo{author}{Teller, E.} (\bibinfo{year}{1953}).
\newblock \bibinfo{title}{{Equation of state calculations by fast computing
  machines}}.
\newblock {\it \bibinfo{journal}{The journal of chemical physics}\/},  {\it
  \bibinfo{volume}{21}\/}, \bibinfo{pages}{1087}.
\bibitem[{Nadler et~al.(2006)Nadler, Lafon, Coifman \& Kevrekidis}]{coiffp}
\bibinfo{author}{Nadler, B.}, \bibinfo{author}{Lafon, S.},
  \bibinfo{author}{Coifman, R.}, \& \bibinfo{author}{Kevrekidis, I.~G.}
  (\bibinfo{year}{2006}).
\newblock \bibinfo{title}{Diffusion maps, spectral clustering and
  eigenfunctions of fokker-planck operators}.
\newblock {\it \bibinfo{journal}{Advances in Neural Information Processing
  Systems}\/},  {\it \bibinfo{volume}{18}\/}, \bibinfo{pages}{955}.
\bibitem[{Neumaier(2001)}]{lsbook}
\bibinfo{author}{Neumaier, A.} (\bibinfo{year}{2001}).
\newblock \bibinfo{title}{{Generalized Lyapunov-Schmidt reduction for
  parametrized equations at near singular points}}.
\newblock {\it \bibinfo{journal}{Linear Algebra and its Applications}\/},  {\it
  \bibinfo{volume}{324}\/}, \bibinfo{pages}{119--131}.
\bibitem[{Papoulis(1977)}]{matchFilter}
\bibinfo{author}{Papoulis, A.} (\bibinfo{year}{1977}).
\newblock {\it \bibinfo{title}{{Signal analysis}}\/}.
\newblock \bibinfo{publisher}{McGraw-Hill New York}.
\bibitem[{Rokhlin \& Tygert(2006)}]{sph_fast}
\bibinfo{author}{Rokhlin, V.}, \& \bibinfo{author}{Tygert, M.}
  (\bibinfo{year}{2006}).
\newblock \bibinfo{title}{{Fast Algorithms for Spherical Harmonic Expansions}}.
\newblock {\it \bibinfo{journal}{SIAM Journal on Scientific Computing}\/},
  {\it \bibinfo{volume}{27}\/}, \bibinfo{pages}{1903}.
\bibitem[{Roweis \& Saul(2000)}]{lle}
\bibinfo{author}{Roweis, S.~T.}, \& \bibinfo{author}{Saul, L.~K.}
  (\bibinfo{year}{2000}).
\newblock \bibinfo{title}{{Nonlinear dimensionality reduction by locally linear
  embedding}}.
\newblock {\it \bibinfo{journal}{Science}\/},  {\it \bibinfo{volume}{290}\/},
  \bibinfo{pages}{2323}.
\bibitem[{Rowley \& Marsden(2000)}]{rowley2}
\bibinfo{author}{Rowley, C.~W.}, \& \bibinfo{author}{Marsden, J.~E.}
  (\bibinfo{year}{2000}).
\newblock \bibinfo{title}{{Reconstruction equations and the Karhunen-Loeve
  expansion for systems with symmetry}}.
\newblock {\it \bibinfo{journal}{Physica D: Nonlinear Phenomena}\/},  {\it
  \bibinfo{volume}{142}\/}, \bibinfo{pages}{1--19}.
\bibitem[{Siettos et~al.(2003)Siettos, Graham \& Kevrekidis}]{LCP}
\bibinfo{author}{Siettos, C.~I.}, \bibinfo{author}{Graham, M.~D.}, \&
  \bibinfo{author}{Kevrekidis, I.~G.} (\bibinfo{year}{2003}).
\newblock \bibinfo{title}{{Coarse Brownian dynamics for nematic liquid
  crystals: Bifurcation, projective integration, and control via stochastic
  simulation}}.
\newblock {\it \bibinfo{journal}{The Journal of Chemical Physics}\/},  {\it
  \bibinfo{volume}{118}\/}, \bibinfo{pages}{10149}.
\bibitem[{Singer(2011)}]{amit}
\bibinfo{author}{Singer, A.} (\bibinfo{year}{2011}).
\newblock \bibinfo{title}{{Angular Synchronization by Eigenvectors and
  Semidefinite Programming}}.
\newblock {\it \bibinfo{journal}{Applied and Computational Harmonic
  Analysis}\/},  {\it \bibinfo{volume}{30}\/}, \bibinfo{pages}{20--36}.
\bibitem[{Singer \& Shkolnisky()}]{cryo2}
\bibinfo{author}{Singer, A.}, \& \bibinfo{author}{Shkolnisky, Y.} ().
\newblock \bibinfo{title}{{Three-Dimensional Structure Determination from
  Common Lines in Cryo-EM by Eigenvectors and Semidefinite Programming}}.
\newblock {\it \bibinfo{journal}{accepted by SIAM Journal on Imaging
  Sciences}\/}, .
\bibitem[{Singer \& Wu()}]{vdm}
\bibinfo{author}{Singer, A.}, \& \bibinfo{author}{Wu, H.~T.} ().
\newblock \bibinfo{title}{{Vector Diffusion Maps and the Connection
  Laplacian}}.
\bibitem[{Singer et~al.()Singer, Zhao, Shkolnisky \& Hadani}]{cryo1}
\bibinfo{author}{Singer, A.}, \bibinfo{author}{Zhao, Z.},
  \bibinfo{author}{Shkolnisky, Y.}, \& \bibinfo{author}{Hadani, R.} ().
\newblock \bibinfo{title}{{Viewing Angle Classification of Cryo-Electron
  Microscopy Images using Eigenvectors}}.
\newblock {\it \bibinfo{journal}{submitted}\/}, .
\bibitem[{Sirisup et~al.(2005)Sirisup, Karniadakis, Xiu \&
  Kevrekidis}]{sirisup}
\bibinfo{author}{Sirisup, S.}, \bibinfo{author}{Karniadakis, G.~E.},
  \bibinfo{author}{Xiu, D.}, \& \bibinfo{author}{Kevrekidis, I.~G.}
  (\bibinfo{year}{2005}).
\newblock \bibinfo{title}{{Equation-free/Galerkin-free POD-assisted computation
  of incompressible flows}}.
\newblock {\it \bibinfo{journal}{Journal of Computational Physics}\/},  {\it
  \bibinfo{volume}{207}\/}, \bibinfo{pages}{568--587}.
\bibitem[{Sirovich(1987)}]{fourOpt}
\bibinfo{author}{Sirovich, L.} (\bibinfo{year}{1987}).
\newblock \bibinfo{title}{{Turbulence and the dynamics of coherent structures.
  I-III}}.
\newblock {\it \bibinfo{journal}{Quarterly of applied mathematics}\/},  {\it
  \bibinfo{volume}{45}\/}, \bibinfo{pages}{561--571}.
\bibitem[{Sivashinsky(1977)}]{ksflame}
\bibinfo{author}{Sivashinsky, G.~I.} (\bibinfo{year}{1977}).
\newblock \bibinfo{title}{{Nonlinear analysis of hydrodynamic instability in
  laminar flames--I. Derivation of basic equations}}.
\newblock {\it \bibinfo{journal}{Acta Astronautica}\/},  {\it
  \bibinfo{volume}{4}\/}, \bibinfo{pages}{1177--1206}.
\bibitem[{Sonday et~al.(2009)Sonday, Haataja \& Kevrekidis}]{physreve}
\bibinfo{author}{Sonday, B.~E.}, \bibinfo{author}{Haataja, M.}, \&
  \bibinfo{author}{Kevrekidis, I.~G.} (\bibinfo{year}{2009}).
\newblock \bibinfo{title}{{Coarse-graining the dynamics of a driven interface
  in the presence of mobile impurities: Effective description via diffusion
  maps}}.
\newblock {\it \bibinfo{journal}{Physical Review E}\/},  {\it
  \bibinfo{volume}{80}\/}, \bibinfo{pages}{31102}.
\bibitem[{Tenenbaum et~al.(2000)Tenenbaum, Silva \& Langford}]{isomap}
\bibinfo{author}{Tenenbaum, J.~B.}, \bibinfo{author}{Silva, V.}, \&
  \bibinfo{author}{Langford, J.~C.} (\bibinfo{year}{2000}).
\newblock \bibinfo{title}{{A global geometric framework for nonlinear
  dimensionality reduction}}.
\newblock {\it \bibinfo{journal}{Science}\/},  {\it \bibinfo{volume}{290}\/},
  \bibinfo{pages}{2319}.
\bibitem[{Titi(1990)}]{titi1990approximate}
\bibinfo{author}{Titi, E.~S.} (\bibinfo{year}{1990}).
\newblock \bibinfo{title}{{On approximate inertial manifolds to the
  Navier-Stokes equations}}.
\newblock {\it \bibinfo{journal}{Journal of mathematical analysis and
  applications}\/},  {\it \bibinfo{volume}{149}\/}, \bibinfo{pages}{540--557}.

\end{thebibliography}
\phantom{blah\\} \noindent{\textbf{Appendix:  initialization of a
probability
distribution with the Metropolis-Hastings algorithm}} \\

To initialize $N$ particles $\{ \mathbf{w}_i \}_{i=1}^N$ on the unit
sphere consistently with some $\psi(\mathbf{u})$, we use the
Metropolis-Hastings algorithm (\cite{metropolis}).
This algorithm may
be used to design a Markov chain with stationary distribution equal
to the desired $\psi(\mathbf{u})$.
After an initial  ``relaxation'' period
of a few iterations, consecutive states $\mathbf{w}_k$ of the chain
are statistically equivalent to samples drawn from
$\psi(\mathbf{u})$.

An auxiliary distribution $q( \bullet | \mathbf{u})$, for
example, a multivariate normal distribution with some mean vector
and covariance matrix, is first selected.
This $q$ distribution is used to generate, from the current state
$\mathbf{w}_k$, a potential next state $\mathbf{w}_{cand}$.
$q$ may be tuned carefully to reduce the variance in the empirically
observed stationary distribution of the Markov chain; for our
purposes, we choose to keep things simple and use $q=1$, meaning
that at each step, we randomly generate a point $\mathbf{w}_{cand}$
on the unit sphere with no regard to the point $\mathbf{w}_k$ from
which it originated.
A candidate state $\mathbf{w}_{cand}$ generated by the auxiliary
distribution is accepted with probability
\begin{equation}
p(\mathbf{w}_k,\mathbf{w}_{cand})= \min \left[1,
\frac{\psi(\mathbf{w}_{cand}) q(\mathbf{w}_k |
\mathbf{w}_{cand})}{\psi(\mathbf{w}_k)q(\mathbf{w}_{cand} |
\mathbf{w}_k)}\right].
\end{equation}
If the candidate $\mathbf{w}_{cand}$ is accepted, the next state
becomes $\mathbf{w}_{k+1}=\mathbf{w}_{cand}$, otherwise if
$\mathbf{w}_{cand}$ is rejected, the next state remains the same as the
current state $\mathbf{w}_{k+1}=\mathbf{w}_k$.
After running the
Metropolis-Hastings algorithm for a large number of iterations, we
subsample the Markov chain to reduce it to $N$ particles $\{
\mathbf{w}_i \}_{i=1}^N$ on the unit sphere.
These $N$ particles
become our consistent initialization according to
$\psi(\mathbf{u})$.

\end{document}